\documentclass[
bibnotes,amsmath,amssymb,aps,
rmp,
]{revtex4-2}

\usepackage{graphicx}
\usepackage{dcolumn}
\usepackage{bm}
\usepackage{hyperref}

\begin{document}

\preprint{APS/123-QED}

\title{Planetary Radar at the Arecibo Observatory}

\author{Michael C. Nolan}
 \email{mcn1@arizona.edu}
\author{Lynn M. Carter}
 \email{lmcarter@arizona.edu}
 \affiliation{Lunar and Planetary Laboratory, University of Arizona}

\author{Edgard G. Rivera-Valent{\'i}n}%
\email{Edgard.Rivera-Valentin@jhuapl.edu}

\affiliation{
 Johns Hopkins Applied Physics Laboratory
}%


\date{\today}

\begin{abstract}
In the late 1990s, the Arecibo Observatory and its planetary radar system were upgraded to increase sensitivity by a factor of 20. This upgrade substantially improved the quality of the data and the ability to observe terrestrial planets, outer planet satellites, planetary rings, and near-Earth objects until the telescope's collapse in 2020. The higher sensitivity allowed radar observations of 889 near-Earth asteroids and comets from 1997 to 2020, compared to the 40 achieved in the previous 30 years, and showed that the population of near-Earth asteroids is heterogeneous, suggesting a wide variety of formation and evolution mechanisms.  The planetary radar's ability to see through the atmospheres of Venus and Titan, into the shadows of Mercury and the Moon, and under the surface of the Moon and Mars provided a unique perspective on those bodies that has driven in-situ exploration. No other existing or planned facility matches the sensitivity that Arecibo had.
\vskip 1 truein
{\large This article may be downloaded for personal use only. Any other use requires prior permission of the author and AIP Publishing. This article appeared in Reviews of Modern Physics 98, 011003 (2026) and may be found at (doi.org/10.1103/k13g-z9s8)}

\end{abstract}

\maketitle
\newpage
\tableofcontents
\section{Introduction}
In the late 1990s, the 305-m William E. Gordon Arecibo Radio Telescope at the Arecibo Observatory, Puerto Rico, underwent the so-called "Gregorian upgrade"  \citep{altschuler_arecibo_2013,campbell_2024}.  This upgrade, which involved adding a Gregorian subreflector to the telescope, improved radar performance by a factor of about 20 due to dramatically decreased thermal noise and more consistent performance as targets transited the Arecibo sky \citep{giovanelli_gregorian_1988}. As a result, it was possible to perform much more systematic studies of planetary targets than before. From 2000 until the telescope's collapse and decommissioning in 2020 \citep{altschuler_arecibo_2013, campbell_2024}, the Arecibo planetary radar was more sensitive than the next-best individual system (the Goldstone 70-meter antenna) by an order of magnitude, and more than the best combined bistatic system (the Goldstone 70-m antenna transmitting to the Green Bank Telescope) by about a factor of two. 

This article presents highlights of the Arecibo planetary radar program, with an emphasis on discoveries enabled by the Gregorian upgrade. We begin with background about the technique of planetary radar and the evolution of the Arecibo facility.  We then describe the key scientific advances enabled by the Arecibo planetary radar system, including some early findings but focusing primarily on the post-Gregorian era. We cover the terrestrial planets, our Moon, and the Saturn system in order of increasing distance from the Sun. Last, we return to the inner solar system, covering near-Earth asteroids and comets, which were the most common planetary radar targets from 2000 to 2020.

\subsection{Planetary Radar}
Radar is a remote sensing technique that uses a transmitter to generate electromagnetic waves, an antenna to aim them at a target, another (or the same) antenna to collect reflections from the target, and a receiver to record the reflected signal. Because the user has control over the transmitted signal, it is possible to learn things about the target that are challenging if not impossible with an incoherent source of radiation (like the Sun). \citet{butrica_see_1996} provides a high-level overview of the state of planetary radar at about the time of Arecibo's Gregorian upgrade. 

The term radar was originally an acronym for RAdio Detection And Ranging. The technique uses radio waves, typically at frequencies of 30-15,000 MHz, corresponding to wavelengths of 10 m to 2 cm. For historical reasons, these have been divided into ``bands'' identified by a letter. Most of the observations performed at Arecibo were at 2380 MHz, 12.6 cm (S-band) or 430 MHz, 70 cm (P-band), with some observations using the 8561 MHz, 3.5 cm (X-band) transmitter on the Goldstone 70-m antenna in California. 

Radio waves can penetrate atmospheres, allowing observations of Venus and Titan. Arecibo's 2380 MHz radar was particularly well suited for Venus observations because higher frequencies suffer from significant atmospheric attenuation \citep{evans_radar_1966}. Like all electromagnetic radiation, radio waves reflect off the surface of a body and also penetrates into nonconductive surfaces, where the penetration depth depends on the composition and mechanical properties of the material and is typically a few wavelengths. This penetration provides information on the subsurfaces of planets, moons, and asteroids. Because radar uses its own illumination source, we can see into shadows on the Moon and Mercury. 

Radar directly measures line-of-sight distance, via the time required for a radio signal to travel to the target and back, and line-of-sight velocity, via Doppler shifting of the frequency of a transmitted signal. The targets observed with the Arecibo planetary radar were, of course, much farther away than is usual for radar on Earth. As a result, it was necessary to use an extremely powerful radar transmitter---one Megawatt of continuous power, formed into a narrow beam by what was effectively the largest radio telescope in the world\footnote{The Jicamarca Radio Telescope is larger, but cannot track astronomical sources \citep{kesaraju_range-doppler_2017}. The larger FAST radio telescope in China became operational in 2020, right at the end of Arecibo's life, but cannot carry the weight of a radar transmitter and its associated cooling}, giving an effective isotropic radiated power of 20 Terawatts at 2380 MHz. For efficiency, rather than use a narrow pulse with the transmitter turned off between pulses, the Arecibo telescope usually transmitted continuously, and the transmitted signal was tagged with a pseudonoise code designed to ``label'' the signal to measure distance. After transmitting for the time required for the radar power to get to a given target and back, the transmitter would be turned off, and then for that same amount of time the signal would be received by the most sensitive radio telescope in the world---Arecibo again. This pattern was usually repeated multiple times over the course of a few hours. \citet{OstroRMP} provides a clear description of planetary radar techniques and a history of results through 1993, shortly before the Gregorian upgrade began.

With care and sufficient signal-to-noise ratio (SNR), it is possible to separately determine the distance (time delay) and speed (Doppler shift) of different points on a target, forming a two-dimensional image of an object, with a spatial resolution that can be much smaller than the optical diffraction limit of both the radar telescope itself and optical telescopes on Earth. As a result, the Arecibo radar data have produced ``spacecraft-flyby-quality'' images of hundreds of objects, obtained at much lower cost than spacecraft observations. Radar observations often provide useful shape information in a short interval---one or a few days of observations, which is particularly important for near-Earth asteroids because they are often only visible for a short period of time. 

These delay-Doppler images suffer from a ``north-south" ambiguity, because generally there are points in the northern and southern hemispheres that map to the same delay and Doppler shift. There are a few ways to mitigate this problem, depending on the target. For the Moon, the radar beam size is smaller than the target, so it is possible to aim the radar at one hemisphere to largely avoid the ambiguity, though there is some leakage through the diffraction lobes of the telescope beam. The early imaging of Venus used interferometry, whereas later observations pointed somewhat away from the planet to emphasize one hemisphere or the other (described in section \ref{sec:Venus}). Mars imaging worked by projecting the images onto a sphere and summing multiple images. Features from the ``wrong" hemisphere smear out and are de-emphasized, and for Mars, the interesting features were often in the north with the corresponding southern features being bland to radar so that they did not dramatically affect the images \citep{Harmon.2012}. For very small targets, such as near-Earth asteroids with unknown shapes, none of these techniques work. If observations are taken from multiple sub-Earth latitudes, typically on different days, different points on the surface have the same delay and Doppler shift. It is possible to model the shape as an inverse problem to resolve the ambiguity, though there are often artifacts \citep{hudson_shape_1994, Nolan2013}. Details of these procedures are available in \citet{OstroRMP} and \citet{magri_radar_2007}.

Planetary radar observations typically transmit a single circular polarization and receive two orthogonal polarizations (either circular or linear). Based on terrestrial and lunar studies, the relative partitioning of the received signal into ``same as transmitted” (SC) and ``opposite of transmitted” (OC) is expected to be determined by surface roughness, with a flat, smooth, perfectly reflecting target giving pure OC reflection. The usual metric for this ``depolarization" from roughness is the ratio of the powers received: Circular Polarization Ratio $\textrm{CPR} = \textrm{SC} / \textrm{OC}$. It is possible to examine more detailed scattering properties, which can be generalized as Stokes parameters describing the total reflectivity, how much the scattering has preserved any polarization, how much of the orthogonal polarization has been produced, and how much the scattering has delayed one polarization with respect to the other. These parameters have been used more often for larger targets, such as the Moon and Venus, because smaller targets often do not provide enough SNR to measure them.

\subsection{Observatory History}
For a complete history of the Arecibo Observatory, we refer the reader to \citet{campbell_2024}, which provides an extensive description of the Observatory's construction, operations, and discoveries. In addition, \citet{altschuler_arecibo_2013} present a short history of the Observatory, written on the occasion of its 50th anniversary.

The Arecibo Observatory was, unlike most facilities, designed to be useful for three different scientific communities. The original idea was for a radar specifically to study the Earth’s ionosphere, but this was modified during the design phase; limited steerability was incorporated to make the telescope useful for astronomical and planetary observations as well \citep{cohen_genesis_2009,campbell_2024}. When the telescope was upgraded in the mid-1970s and again in the late-1990s Gregorian upgrade, the latter fields were the primary beneficiaries. 

The original Arecibo telescope used a spherical primary mirror combined with ``line feed" antennas that corrected the extreme spherical aberration of the primary mirror, but only worked over a narrow range of frequencies. The Gregorian upgrade replaced most of these tuned line feeds with a pair of aberration-correcting mirrors that worked over a wide range of frequencies and directed radio signals at the primary mirror over a broader span of pointing directions than did the line feeds. The sensitivity of a planetary radar system is described by the ratio of the received power to the variations in receiver noise power. The latter is largely thermal noise, described by a characteristic temperature $T_\mathrm{sys}$: SNR $\propto P_\mathrm{rcv}/T_\mathrm{sys}$. For a point-source target (for Arecibo, all of the planetary targets except the Moon), that  is \citep[after][eq. 1]{OstroRMP}:
\begin{equation}
    \mathrm{SNR} \propto P_\mathrm{tx}G_\mathrm{tx}G_\mathrm{rcv}\lambda^2/R^4 T_\mathrm{sys}
\end{equation}
where $P_\mathrm{tx}$ is the transmitter output power, $G_\mathrm{tx}$ is the telescope gain while transmitting, $G_\mathrm{rcv}$ is the telescope gain while receiving, $\lambda$ is the wavelength of the radio waves, and $R$ is the distance to the target.

The Gregorian upgrade to the primary mirror, the suspended secondary mirror \citep{giovanelli_gregorian_1988,kildal_arecibo_1994}, and the antenna feeds, as well as the radar transmitter and receiver, brought a factor of 20 improvement in sensitivity (Table~\ref{tab:AOprop}). Except for transmitter power, which was doubled, the Gregorian upgrade only slightly improved the peak values of the telescope parameters, that is, when the telescope was pointed directly at the zenith. However, the pre-upgrade telescope performance fell off dramatically as the telescope was pointed away from zenith as the target moved across the sky, with $T_\mathrm{sys}$ doubling and $G_\mathrm{tx}$ and $G_\mathrm{rcv}$ both decreasing by about 1/3 \citep{giovanelli_gregorian_1988}. All of these factors were significantly improved by the upgrade so that the performance remained approximately constant from 0 to 15$^\circ$ zenith angle and degraded more slowly to the maximum 20$^\circ$ zenith angle of the telescope.

\begin{figure}
\includegraphics[width=0.32\linewidth]{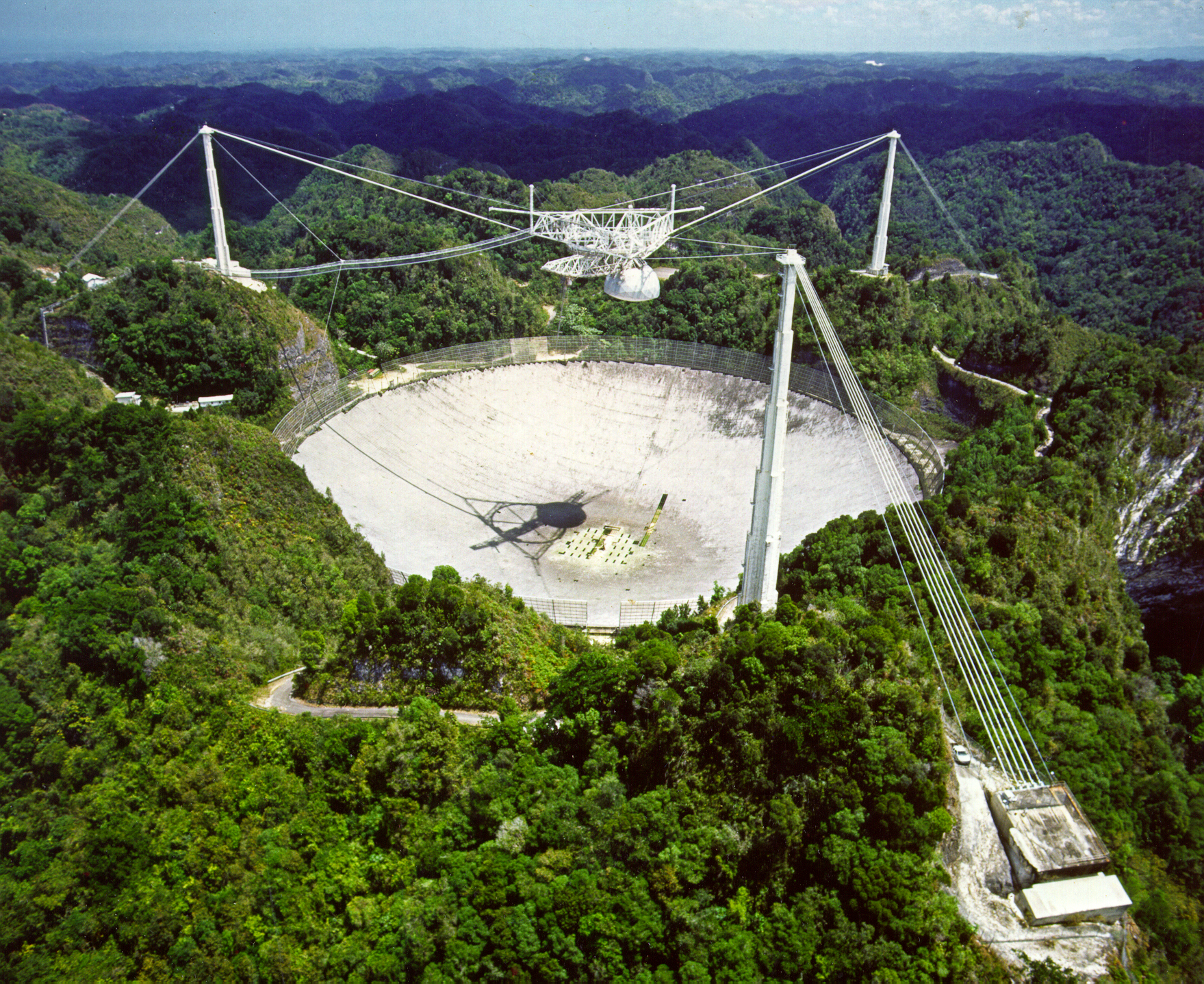}
\includegraphics[width=0.4\linewidth]{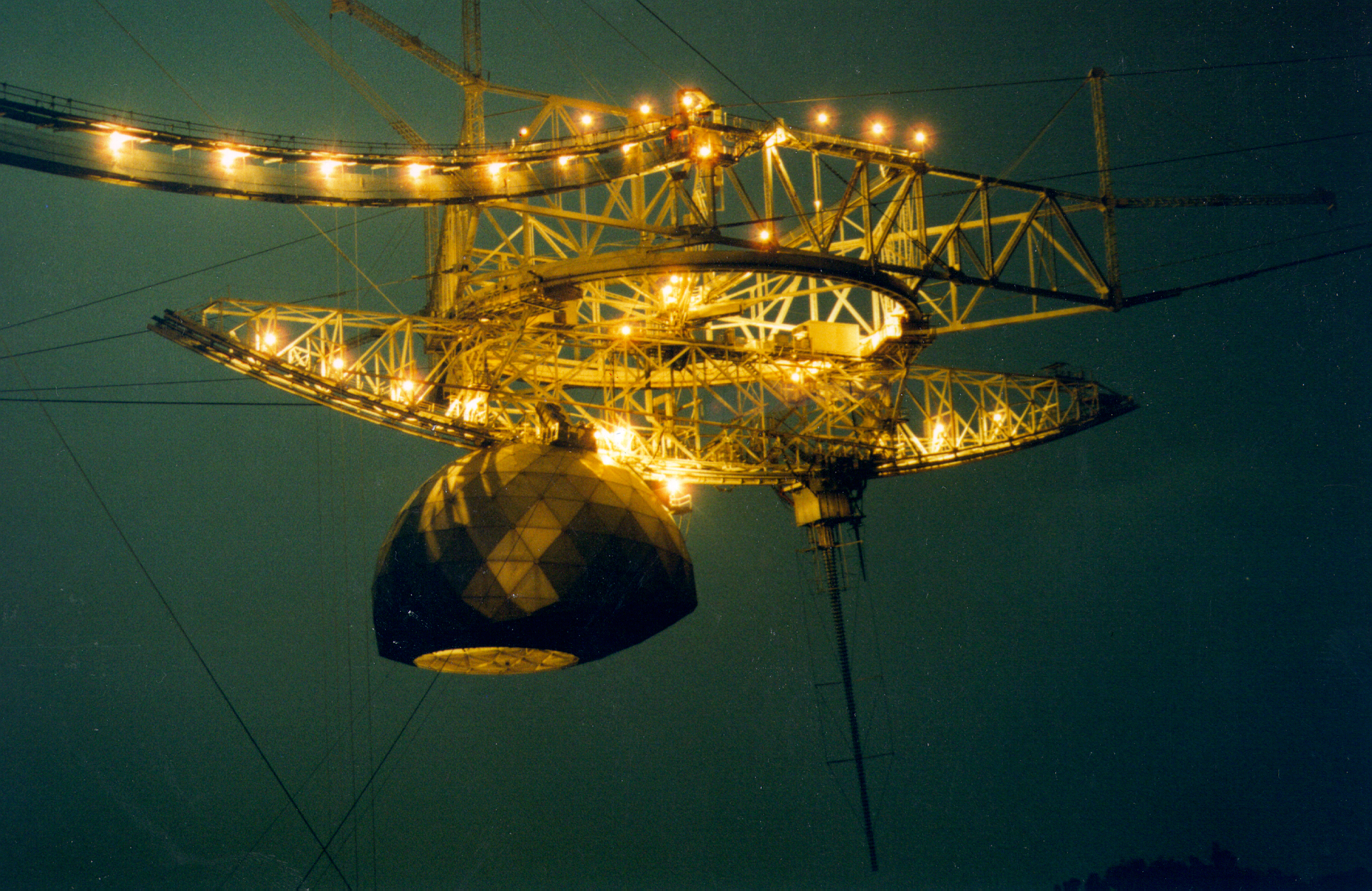}
\includegraphics[width=0.18\linewidth]{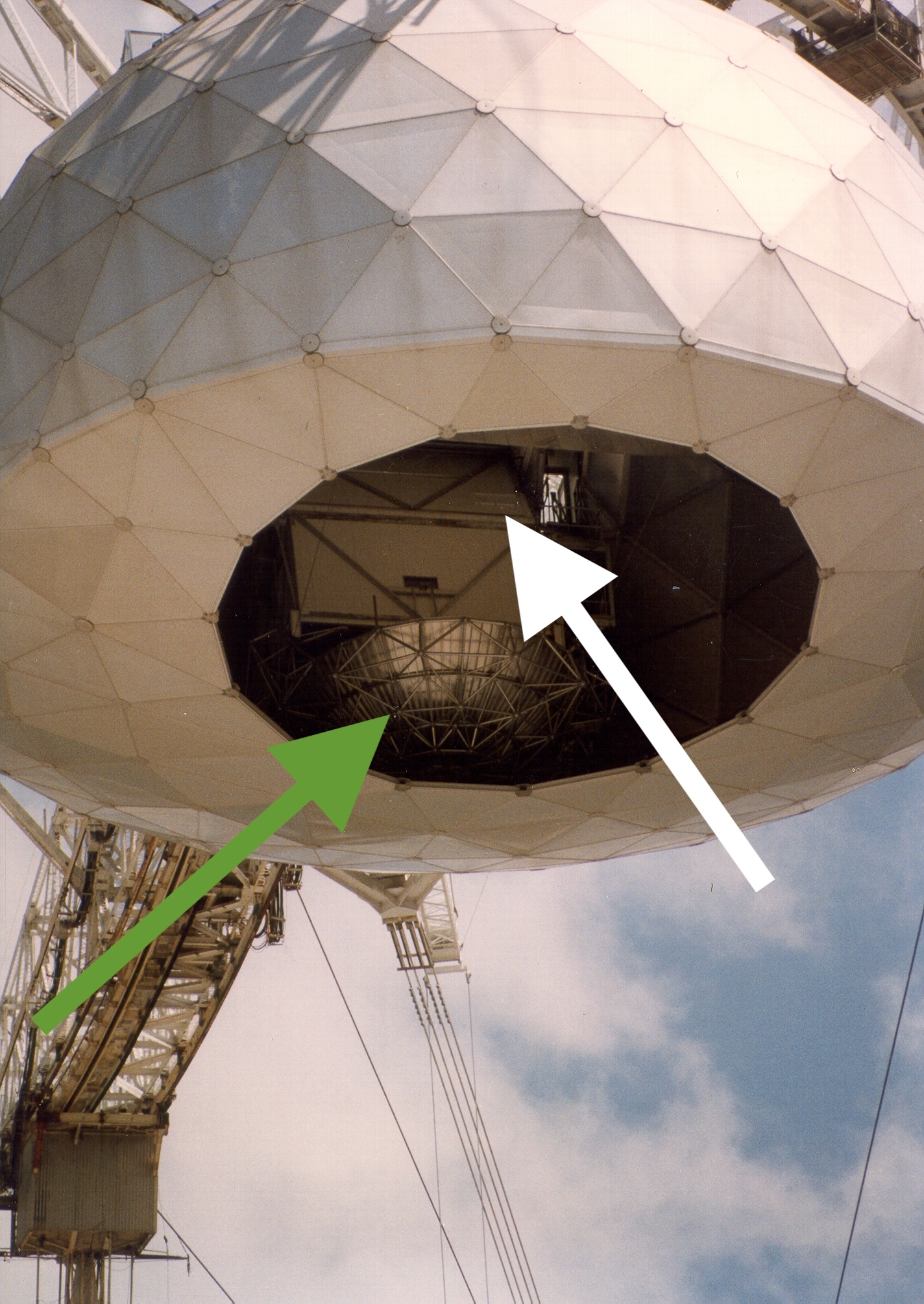}
\caption{\label{fig:AO}The left image shows the Arecibo telescope. The 305-m diameter primary reflector consists of 38778 aluminum panels suspended on cables to form a spherical surface. The secondary structure was suspended 150 m above the dish on steel cables running from three concrete towers. The center image shows the Gregorian subreflector in its radome and the one remaining line feed used for the 430-MHz radar. They could move along the curved (banana-shaped) rails to select the telescope pointing in elevation. The elevation rail rotated on the visible circular azimuth ring. The right image shows the underside of the tertiary mirror inside the radome aperture (green arrow) with the ``receiver room'' (white arrow) above it. The small white mark visible near the top of the aperture just above the white arrowhead is a standard-sized door for scale. Radio waves from the sky reflected from the spherical primary mirror and entered the radome through the aperture, reflected again from the 27-m secondary mirror (not visible) that covered most of the top of the radome, and reflected yet again from the tertiary mirror into feed horns that pointed down from a lazy Susan under the receiver room. When transmitting, the 1-MW signal transited that path in the opposite direction. Image credit: Courtesy of the NAIC -- Arecibo Observatory, a facility of the NSF. Photos by Tony Acevedo.}
\end{figure}

Figure~\ref{fig:AO} shows the telescope as it appeared in the early 2000s. The secondary and tertiary mirrors in the Gregorian radome were highly curved in order to correct the spherical aberration of the primary mirror in a relatively small space with an aperture efficiency of about 40\%.

\begin{table}\begin{tabular}{l r}
Telescope Geodetic Latitude (GRS80)&18.344219 deg\\
Telescope Geodetic Longitude (GRS80)&293.247306 deg\\
Telescope Altitude (GRS80)&453.34 m\\
Elevation Range&70.31--88.94 deg\\
Transmitter Operating Center Frequency&2380 MHz\\
 Transmitter Final Amplifier Bandwidth&27 MHz 1-dB\\
 Transmitter Output Power&1 MW continuous\\
 System Output Power&20 TW Effective Isotropic Radiated Power\\
 Telescope Gain (transmitting)& 73.4 dB\\
 Telescope Effective Area (receiving)& 27,610 m$^2$\\
 Receiver System Temperature&23--30K\\
 Narrowest Transmitter Baud Length&50 ns, 20 MHz\\
 Finest Range Resolution&7.5 m
 \end{tabular}
     
     \caption{\label{tab:AOprop}Basic system properties of the Arecibo Observatory planetary radar system from 1997 to 2020, after the Gregorian upgrade. There were several intervals where the performance was reduced due to problems with various components, all of which were custom-built for the Observatory. Position uncertainty is approximately 1 m. Altitude is the location where the elevation and azimuth axes intersect, above the telescope structure. Pre-upgrade parameters depended strongly on telescope pointing so that the transmitter gain multiplied by the telescope effective area divided by the system temperature integrated across the track of a target as it moved through the sky was approximately 10\% of the values given here. The transmitter power was also only 0.5 MW, resulting in a factor of 20 improvement altogether.}
\end{table}

The earliest planetary observations with the Arecibo telescope in the 1960s concentrated on Venus, Mercury, and the Moon. Mars was also observed beginning in 1963, but not imaged at Arecibo until 1990. Because Arecibo can only observe objects north of the equator, Saturn's orbit period of $\sim 30$ y meant that the first published observations (of the rings) were not made until 1976 \citep{pettengill_saturns_1976,goldstein_rings_1977}.

With the Gregorian upgrade, asteroid astronomy was seen as the likely driver of planetary radar at Arecibo \citep[][p 261]{butrica_see_1996},\footnote{Michael Nolan is the ``planetary astronomer with an interest in asteroids" referred to in that section, then about to be hired as a post-doctoral researcher at the Arecibo Observatory, where he remained for twenty years.} owing to the recently acquired understanding of the potential hazards posed by asteroids in near-Earth space. The upgrade occurred just as automated asteroid search programs increased the near-Earth asteroid discovery rate to more than 100 per year in 1998\footnote{The discovery rate had increased to thousands per year by 2020.} \citep{stokes_lincoln_2000,jedicke_observational_2002,jedicke_surveys_2015}. At the same time, Saturn was at its most northern aspect as viewed from Earth at opposition. As a result, the Arecibo planetary radar program was well situated to begin observing these objects, continuing (with some hiccups) until the collapse of the telescope in 2020.

Even then, though, there was concern about ``budget cuts in NASA and U.S. scientific research in general" \citep{butrica_see_1996}. These cuts began to be realized at Arecibo after a freak accident killed J{\"u}rgen Rahe, NASA's Director for Solar System Exploration, in 1997 \citep{Pilcher1997Jurgen}\footnote{A storm knocked over a tree, which fell on his car. He was the storm's only known casualty.}.
In the resulting reorganization, NASA stopped supporting ground-based facilities not directly tied to spacecraft mission support. Thus, NASA began reducing its support of the Arecibo planetary radar program in 2002 and eliminated it in 2004 \citep{feder_planetary_2003,campbell_2024}. 

One of the justifications NASA gave was that the budget of the U.S. National Science Foundation (NSF), the federal agency with direct oversight of the Observatory, was ``growing like a bandit" \citep{feder_planetary_2003}, with a promised doubling of its budget \citep{showstack_bill_2002} to begin new projects. However, NSF's budget actually remained flat, leaving the agency struggling to complete those projects. Thus, it was announced in 2006 that the Observatory would be decommissioned if outside funding could not be found \citep{brumfiel_astronomers_2006,kumar_panel_2006,blandford_ground_2006}.

Fortunately, around this same time, interest in near-Earth asteroids was increasing. In 2008, bills to support the Arecibo planetary radar program were introduced in the U.S. Congress, and \$2 million was appropriated in the budget of fiscal year 2010. This funding helped the Arecibo Observatory continue to operate and to expand its program of planetary radar observations. By 2012, the program constituted approximately 1/3 of the Observatory budget, and it remained strong until the failure of the telescope in 2020, observing tens to $>100$ objects in a typical year \citep{virkki_arecibo_2022}.

In spite of the budget difficulties, the new capabilities provided by the dramatically improved sensitivity and resolution of the upgraded Arecibo planetary radar led to major discoveries and advances in the field, as we describe in the following sections. 

\section{Mercury}
Observations of Mercury were among the earliest results at the Observatory. Before 1965, Mercury was believed to rotate in one Mercurian year, or 88 days, as measured by \citet{schiaparelli_sulla_1890}. \citet{pettengill_dyce_1965} used delay-Doppler measurements from Arecibo to enhance the visibility of the edges of the spectrum and showed that the actual rotation rate was significantly faster than that---$59 \pm 5$---days, consistent with three rotations every two Mercurian years. Over the next few years, a theoretical basis for this rotation rate was developed, which relies on the large eccentricity of Mercury's orbit \citep{colombo_rotation_1966}.

Also prior to the Gregorian upgrade at Arecibo, full-disk monostatic observations of Mercury using the Arecibo S-band transmitter, together with bistatic measurements using the Goldstone X-band transmitter and 26 antennas of the Very Large Array as receivers, revealed radar-bright features at the poles \citep{Harmon.1992, Slade.1992}. The brightness and high circular polarization ratio of these features were similar to the properties of the icy moons of Jupiter and the layered deposits at the Martian poles \citep{Campbell.1977, Ostro.1980, Ostro.1990, Muhleman.1991}. Thus, the features were considered likely to be the result of scattering from nearly pure water ice in permanently shadowed regions (PSRs) at Mercury’s poles \citep{hapke_coherent_1990,Butler.1993,black_icy_2001}, which were shown to have thermally favorable environments for ice stability \citep{Ingersoll.1992, Paige.1992}. The coarse resolution of this imagery, on the order of 100~km at the subradar point, did not readily permit tracing of all radar-bright features to specific geologic units, particularly at the north pole. Subsequent Arecibo observations, though, improved the delay and Doppler resolution, and thus the projected spatial resolution, by an order of magnitude. The radar-bright features at both poles were traced to numerous craters \citep{Harmon.1994}. 

The Gregorian upgrade to the Arecibo planetary radar system allowed for another improvement of radar images of Mercury to a spatially projected resolution of 1.5 km \citep{Harmon.2001}. The imagery confirmed previous results and identified additional, smaller radar-bright features extending from the poles to as low as $\sim$70$^{\circ}$ latitude in both hemispheres. Thermal modeling showed that although surface water ice is stable in bowl-shaped craters within 2$^{\circ}$ of Mercury's poles, ice deposits beneath a thin regolith layer can be stable within 10$^{\circ}$. Throughout the early 2000s, an observing campaign at Arecibo gathered images over the full range of subradar longitude aspects and latitudes from -7.5$^{\circ}$ to 11.8$^{\circ}$ \citep{Harmon.2011}. This observing campaign, and the resulting varying viewing geometries, minimized the bias of radar shadowing (where the interior of a crater is not visible to the radar). This improves the ability to distinguish PSRs truly lacking a radar signature, rather than those that were not sampled by the radar, and thus with little to no ice.  Understanding the distribution of water ice at Mercury's poles helps to constrain potential formation hypotheses. 

The discovery of ice at Mercury’s poles partially motivated NASA’s MESSENGER mission, which launched in 2004 and operated at Mercury between 2011 and 2015. The detailed spacecraft observations confirmed that all radar-bright features are associated with PSRs, but that not all PSRs are occupied by a radar-bright reflector \citep{Chabot.2012, Deutsch.2016, Chabot.2018}. In fact, \citet{Chabot.2018} showed that approximately 50\% of Mercurian PSRs are probably ice-free, indicating that the volatiles were likely delivered by an episodic event, such as a cometary impact. However, radar illumination analysis using high-resolution topographic maps generated from MESSENGER spacecraft data suggested that not all Mercurian PSRs were adequately sampled during the Arecibo radar campaign post-upgrade \citep{Glaser.2023}, opening the potential for a larger volatile reservoir. The improved thermal modeling afforded by MESSENGER’s topographic maps also facilitated local-scale analysis of the high-resolution Arecibo radar imagery. For example, \citet{Chabot.2014} showed that PSR boundaries within large craters, such as Prokofiev (85.7$^{\circ}$, 297.1$^{\circ}$W), agree well with the location of radar-bright features. The distinct and clear boundaries within Prokofiev and other craters suggest that the Mercurian ice deposits are geologically young. \citet{Glantzberg.2023} also showed a correlation between radar backscatter intensity and local maximum temperature. Additionally, MESSENGER reflectance measurements showed that although some PSRs are anomalously bright in visible light, indicative of exposed water ice, others are anomalously dark, suggesting burial by a low-reflectance, perhaps organic-rich material \citep{Neumann.2013, Chabot.2014, Chabot.2016, Chabot.2018, Hamill.2020}. Arecibo radar brightness measurements across these features indicated that the dark overburden is probably thin, on the order of a few centimeters. If it was produced by sublimation and subsequent exposure and accumulation of dark constituents within the ice, the thickness of the lag deposit would further constrain the timing of volatile delivery and deposition.  

In 2017, Hurricane Maria devastated Puerto Rico and damaged the Arecibo Observatory, resulting in lower transmit power and reduced gain. Nevertheless, subsequent S-band radar observations of Mercury’s poles continued to provide valuable new insights, particularly when paired with MESSENGER data \citep{Rivera-Valentin.2022}---a testament to the unsurpassed capabilities of the post-upgrade system. \citet{Rivera-Valentin.2022} identified additional local-scale heterogeneities within the northern radar-bright features. Further, using the high-resolution MESSENGER topographic maps \citep{Barker.2022} to calculate the local radar incidence angle \citep{Mazarico.2011}, \citet{Rivera-Valentin.2022} were able to investigate the scattering properties of these features. Using the new radar imagery, they showed that some craters have a central bright feature surrounded by lower backscatter in a gradational pattern, whereas others present a mottled pattern. Their radar scattering models showed that the high radar backscatter regions are consistent with nearly pure water ice, while the lower-reflectivity regions may have $>$20\% impurities. This heterogeneity was confirmed by \citet{Glantzberg.2023}, who showed that environments where surface water ice is stable present significant variation in radar backscatter. Such variation may be attributable to differences in ice purity, abundance, and regolith mantling; these differences may be driven by surface processes over time and/or related to the depositional environment. 

Insights gained from the high-resolution Arecibo radar imagery of Mercury help inform the European Space Agency’s upcoming Mercury Planetary Orbiter (MPO), part of the BepiColombo mission\citep{Benkhoff.2021}. Unlike MESSENGER, MPO's polar orbits will allow for detailed measurements of Mercury’s north pole. Measurements using MPO’s laser altimeter instrument could be made over craters where Arecibo radar imagery has indicated heterogeneity in backscatter, placing further constraints on the ice content, formation, and evolution of PSRs. Considering Mercury's deposits as the canonical example of ice within PSRs, analysis of their radar scattering properties can also further the identification and characterization of potential buried ice within lunar PSRs, which can assist with the crewed exploration of the Moon.

\section{Venus}\label{sec:Venus}
Cloud-covered Venus has long been a target of ground-based radar systems. Early Venus observations contributed to determining its rotation period and surface features. Venus' rotation had been determined to be retrograde and slow by earlier radar measurements at the Jet Propulsion Laboratory's Goldstone facility, Lincoln Laboratory's Millstone Hill, and Jodrell Bank \citep{smith_radar_1963,goldstein_rotation_1963,ponsonby_rotation_1964}, but the higher sensitivity of Arecibo compared to other facilities led to an improved measurement of the rotation rate and pole direction. \citet{dyce_radar_1967} measured the rotation period to be $245.1 \pm 0.7$ days, but they chose to assign an uncertainty of 2 days to account for known but hard-to-quantify systematic errors. The currently accepted value is 243.0 days, so they chose wisely. All of the groups working on this problem noted that the rotation period is nearly the same as the 243.16-day 8:13 resonance period that would present the same face of Venus to Earth on every orbit, and \citet{dyce_radar_1967} do mention this fact, which may have played into their uncertainty analysis. 

The Arecibo upgrades provided an opportunity to obtain new data of Venus, particularly higher-SNR polarimetry, higher-spatial-resolution imaging, and longer-duration surface change detection compared with observations from the 1980s. Because Venus always presents nearly the same side to Earth during close approaches, Venus radar images acquired from Earth usually have a limited range of longitude values, ranging from about -80$^{\circ}$ E (280$^{\circ}$) to +50$^{\circ}$ E. The latitudinal position of the subradar point changes between inferior conjunctions from about 9$^{\circ}$ N to 9$^{\circ}$ S, altering the radar incidence angles and also providing more expansive imaging views of either the northern or southern hemispheres \citep[e.g.,][]{Campbell2022PSJ}. This viewing area includes parts of the southern hemisphere that were not imaged by the Magellan spacecraft \citep{Kratter2007JGRE}.

The beamwidth of the Arecibo telescope at 12.6 cm wavelength is 2 arcminutes, which is about twice the angular size of Venus at inferior conjunction. Because of that, points in the northern and southern hemispheres are mixed together in radar imagery because they are at the same distance and moving at the same speed; this is called the ``north-south ambiguity''. Two different methods were used to reduce the effects of this ambiguity.

In the 1960s, observations with the 70-cm radar, where the beamwidth was even wider than at 12.6 cm, a second antenna (built by an amateur radio enthusiast \citep{campbell_2024}) located 11 km away was used as a radio interferometer, which made it possible to separate the signals from the two hemispheres electrically \citep{campbell_1970}. Observations in the 1970s used this same technique with a purpose-built antenna \citep{campbell_new_1976}. Later observations achieved this separation by pointing the telescope north and south of the planet on alternating observing periods to increase the power for one hemisphere and lessen the effects of the ambiguities \citep{Campbell2022PSJ}.

Over the years from 1999 to 2004, the S-band radar system was used to acquire data with the purpose of generating Stokes polarization parameter maps of the surface of Venus and searching for anomalies that could reveal changes in the surface structure \cite{Carter2006JGRE, Carter2004JGRE, CarterProcIEEE, Kratter2007JGRE}. The radar resolution was selected to be either 4.2 or 8.0 $\mu$s baud length, to maximize the SNR. The data were used to generate maps of all four Stokes polarization parameters, as well as the degree of linear polarization and the circular polarization ratio. Although circular polarization ratio data from Venus had been used before \citep{Campbell1992JGR}, the linear polarization maps were new and revealed correlations with many different types of geologic features.  For the case of a circularly polarized wave incident on a natural rock surface, a linearly polarized echo component is most often generated by penetration into the surface and reflection from buried rocks or structures. The degree of linear polarization maps thus demonstrated that some areas of Venus have mantling material at least $\sim$1 cm thick \citep{CarterProcIEEE}, much of which was difficult to detect by the Magellan spacecraft synthetic aperture radar (SAR) system, which did not have polarimetry. 

\begin{figure}
    \includegraphics{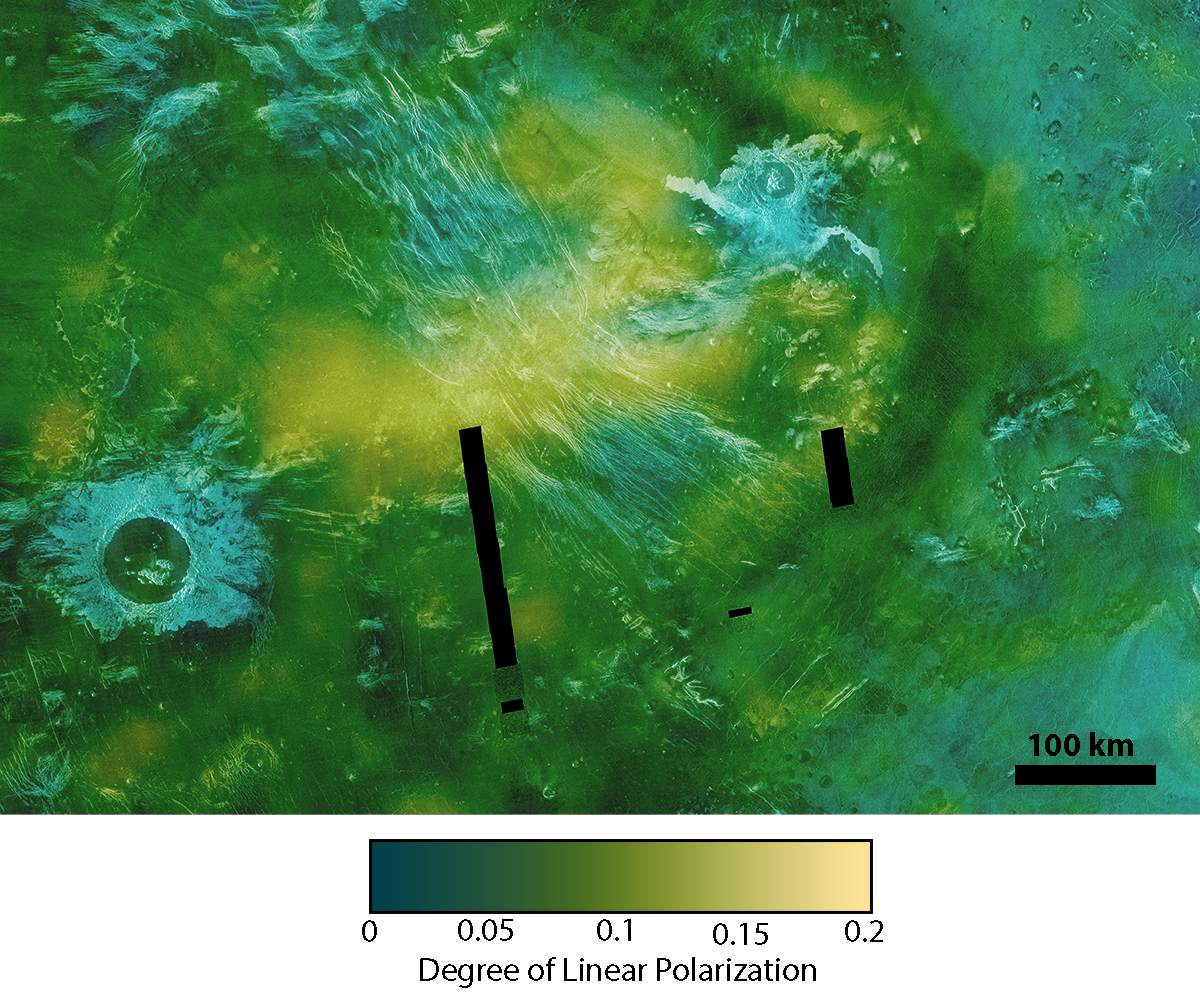}
    \caption{\label{fig:carson} Map of the degree of linear polarization of Carson crater, overlaid on a Magellan SAR image. The map shows that radar-bright areas of the Carson parabola are mantled by impact ejecta. These bright areas are likely a mix of fine-grained material and rugged rocks or perhaps surface dunes. Wind streaks are visible in the upper left---evidence of reworking of the crater ejecta. The pixel resolution of the DLP maps is 12 km. Adapted from \citet{Carter2004JGRE}.}
\end{figure}

Impact crater ejecta, dune fields, wind streaks, and volcanic centers are some of the major features that show evidence of mantling material (Figure~\ref{fig:carson}). Approximately 45 impact craters, including five with large parabola-shaped deposits, exhibit increases of 10--40\% in linear polarization relative to the surrounding areas, suggesting mantling by fine-grained ejecta material \citep{Carter2004JGRE}. For some craters, such as Galina, the radar polarimetry provides evidence that the surrounding radar-dark areas are covered in fine mantling material, instead of being smooth and swept clean as proposed during the Magellan mission. In other cases, such as Carson (Figure~\ref{fig:carson}) and Anya craters, the increased degree of linear polarization is correlated with radar-bright areas, suggesting a more complex surface with both embedded rocks and fine material. Some individual lava flows and dome fields are also associated with increased degrees of linear polarization. It is possible that the radar penetrates the flows and reflects from buried air gaps, a situation that can occur on Earth \citep{Carter2006JGRE}. Alternatively, the flows may also be mantled in thin layers of dust or tephra (pyroclastic deposits). Interestingly, some of the high-reflectivity, low-emissivity summits on Venus (e.g., Theia Mons, Tepev Mons, shield volcanoes located on highland regions in the northern hemisphere) also have patches of increased linear polarization, suggesting that the radar wave is penetrating into some surface mantling material \citep{Carter2006JGRE}.

Higher-resolution imaging data were acquired from 2012 to 2022, in some cases using the Robert C. Byrd Green Bank Telescope as a receiver station, and processed to a $\sim$1 km pixel size \citep{Campbell2022PSJ, Campbell2021pdsdata}.  Data from multiple years with similar viewing geometries can be combined to achieve a higher SNR \citep{Campbell2022PSJ}. Although their data were not processed for Stokes polarimetry, the OC, SC, and CPR maps can be used to look for changes in surface roughness. These images showed that some portions of the radar-bright ``tesserae” (or cratons) have a lower SC-polarized signature attributable to smoothing of the surface by the deposition of crater ejecta. In particular, a large parabola ejecta blanket associated with Stuart crater mantles a large portion of Alpha Regio, the planned landing site of the NASA DAVINCI spacecraft \citep{Campbell2015Icarus}. The 25\% lower SC-polarized echo compared to nearby terrain suggests a mantling deposit thickness of a few centimeters \citep{Campbell2015Icarus}. Other regions may be covered by $>$10 cm of crater ejecta \citep{Campbell2015Icarus}. The distribution and thickness of mantling deposits have implications for how sediments are formed and transported on Venus. 
These higher-resolution Arecibo data were also used to investigate the surface properties of diffuse-margin, radar-bright deposits near the summits of some volcanoes and coronae. These bright features have relatively high CPR values that indicate rugged surfaces or perhaps column-collapse pyroclastic deposits including embedded rocks \citep{Campbell2017JGRE, 2022JGREGanesh}.  The deposits are stratigraphically higher and younger than surrounding features, and they may be evidence of a fairly recent renewal of magmatic activity and tectonics in these areas \citep{Campbell2017JGRE}.

The long duration of Venus observations at Arecibo provided an opportunity to estimate the mean rotation rate by measuring the longitudinal offset of small surface features such as crater central peaks \citep{Campbell2019Icarus}. Multiple datasets acquired between 1988 and 2017 have a range resolution of $\sim$600 m. For determining the rotation rate, the data were not mapped to a cartographic grid because delay-Doppler images are sufficient to detect and measure the position of features and determine the subradar point longitude \citep{Campbell2019Icarus}. These offset measurements were used to compute a length-of-day of $243.0212\pm0.0006\,$d, slightly different from the value of 243.0185$\,$d accepted by the International Astronomical Union. This difference means that the positional offsets from features on the Magellan cartographic grid are already $>$20 km; this information is critical for landing future missions on the surface \citep{Campbell2019Icarus}.

Finally, the recent maps of Venus at 1--2 km spatial resolution  can be compared with similar-resolution maps collected in 1988 to search for surface changes \citep{Campbell2022PSJ}. The 1988 data have lower SNR due to the older Arecibo radar system, and they also have somewhat different viewing geometries, which makes it difficult to search for small changes in images \citep{Campbell2021LPI}. In addition, the north-south ambiguity contributes some spurious power to the opposite hemisphere, so small changes in the reflectivity are also likely inconclusive. Comparison of images from 2017 and 1988 show no apparent surface changes on the scale of 10 km$^{2}$ or larger \citep{Campbell2021LPI}. Although no surface changes have been detected yet, it is still possible that smaller-scale changes have occurred, such as those observed using radar data from the Magellan spacecraft \citep{Herrick.2023}, or that changes have occurred outside of the Arecibo prime viewing area of Venus. The data produced by Arecibo over the past decades have provided a bridge between the Magellan spacecraft mission in the 1990s and the future Venus radar mapping missions such as ESA EnVision and NASA VERITAS that are scheduled to arrive in the 2030s.

\section{Moon}
The Moon is a particularly attractive target for radar remote sensing because its low density regolith is typically very penetrable to radar, and its proximity to Earth enables very high-SNR images at high spatial resolutions. Both the S-band radar system and the longer-wavelength P-band system (430 MHz, 70 cm wavelength) at Arecibo were used to observe the Earth-facing side of the Moon between 2001 and 2015 \citep{Campbell2007pdsdata, Campbell2011pdsdata, Campbell2016PASP}. The round-trip light travel time to the Moon is less than 3 s, so these observations used a bistatic geometry with the Robert C. Byrd Green Bank Telescope as the receiver. P-band observations used either a 3 $\mu$s pulse with a pulse repetition period of 15 ms and a 17-minute coherence interval for a resolution of $\sim$450 m, or a 13-element barker code with a 1-$\mu$s baud length and a coherence length of 40 minutes for a resolution of $\sim$200 m \citep{Campbell2007ITGRS, Campbell2016PASP}.  Some S-band data were collected with a baud length of 0.1 $\mu$s and a coherence interval of $>$50 minutes, which led to single-look images with $\sim$20 m resolution \citep{Campbell2006Nature}. However, this setup required sampling with only 2 bits, which sometimes led to issues with clipping and poor polarimetry performance. Most data were then collected with a 0.2-$\mu$s baud length and 4-bit sampling and processed to a spatial resolution of 80 m with four looks \citep{Campbell2010Icarus, Campbell2016PASP}. The radar system used a predicted Doppler shift to correct the central transmitted frequency to place one selected target near the image center. The data were processed using an autofocusing technique to reduce the smearing caused by the Moon's motion and were mapped to latitude/longitude space \citep{Campbell2007ITGRS, Campbell2010Icarus}. 

\subsection{Lunar Poles}
The possible presence of ice at the lunar poles, and the form and distribution of any ice, has implications for both the evolution of lunar volatiles and future exploration by humans. The discovery of enhanced hydrogen in PSRs in polar craters by the Lunar Prospector mission \citep{1998SciFeldman} motivated the observation of these PSRs with radar, because radar waves can penetrate into the surface and radar backscatter sometimes experiences a coherent backscatter effect that can be diagnostic of ice \citep{harmonMercury, black_icy_2001}. Due to the pattern of the lunar libration, views of the southern hemisphere and south pole PSRs were more optimal than northern views during the period of most of the Arecibo observations.  

\begin{figure}
    \includegraphics{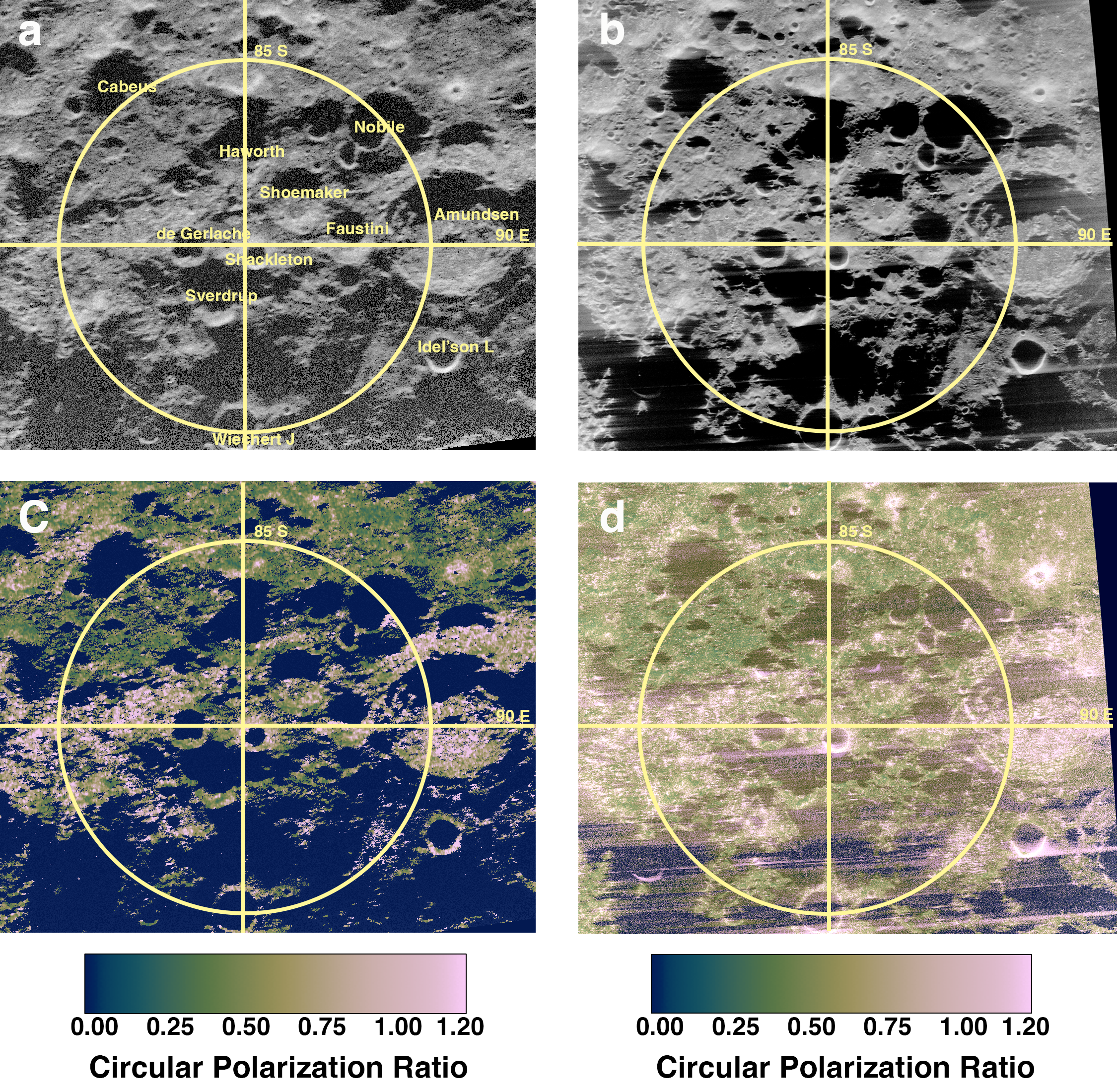}
    \caption{\label{fig:lunarpole} Arecibo radar images of the lunar south pole. (a) P-band SC radar image with craters labeled. (b) S-band SC radar image. The bright, radar-facing crater walls cause image artifacts extending horizontally. (c) P-band CPR image overlaid on the radar backscatter. CPR values $>1$ are observed for some bright crater ejecta, smooth crater floors, and intercrater highlands. (d) S-band CPR image overlaid on the radar backscatter. The processing artifacts obstruct some areas of the image, but overall the CPR values have a similar trend to that in the P-band data.}
\end{figure}

Observations of the lunar south pole with the Arecibo P-band radar system revealed changes in the CPR across the pole (Figure~\ref{fig:lunarpole}) but no significant increase in backscatter power and CPR from the PSRs \citep{Campbell2003Nature, Campbell2006Icarus}. Instead, the increased CPR values were associated with fresh young craters with rocky ejecta, or areas of smooth plains that were deposited during the Orientale impact \citep{Campbell2006Icarus}. The P-band radar can potentially penetrate several meters to several tens of meters into the subsurface, so the lack of a clear coherent backscatter signature in the PSRs suggested that there are no large blocks of pure ice buried at depth.  However, the spatial resolution of 450 m could prevent detection of small patches of ice, and the radar coherent backscatter signature has been observed to be stronger at S-band than P-band frequencies \citep{Black2001Icarus}.

The lunar south pole was observed with the Arecibo S-band radar system at 20 m spatial resolution in 2005 \citep{Campbell2006Nature} (Figure~\ref{fig:lunarpole}) though later observations were performed at 80 m resolution to improve calibration accuracy and SNR \citep{Campbell2010Icarus}. As with the P-band data, the shorter-wavelength data showed no clear evidence of coherent backscatter from surface or near-subsurface ice in PSRs. Bright, young craters and other rocky surfaces like crater walls have high CPR values, but the PSRs have CPRs that are similar to those found in sunlit craters near the pole. This suggested that ice, if present, is likely to be small grains or impure patches \citep{Campbell2006Nature}. The lack of conclusive evidence for ice in the Arecibo datasets places limits on the type and amount of ice that can exist at the poles. The search for buried lunar ice continues, and if significant patches of subsurface ice are eventually found, it will be useful to understand how the nature of the ice contributed to the lack of clear radar signatures.

\subsection{Volcanism}
\begin{figure}
    \includegraphics[scale=0.60]{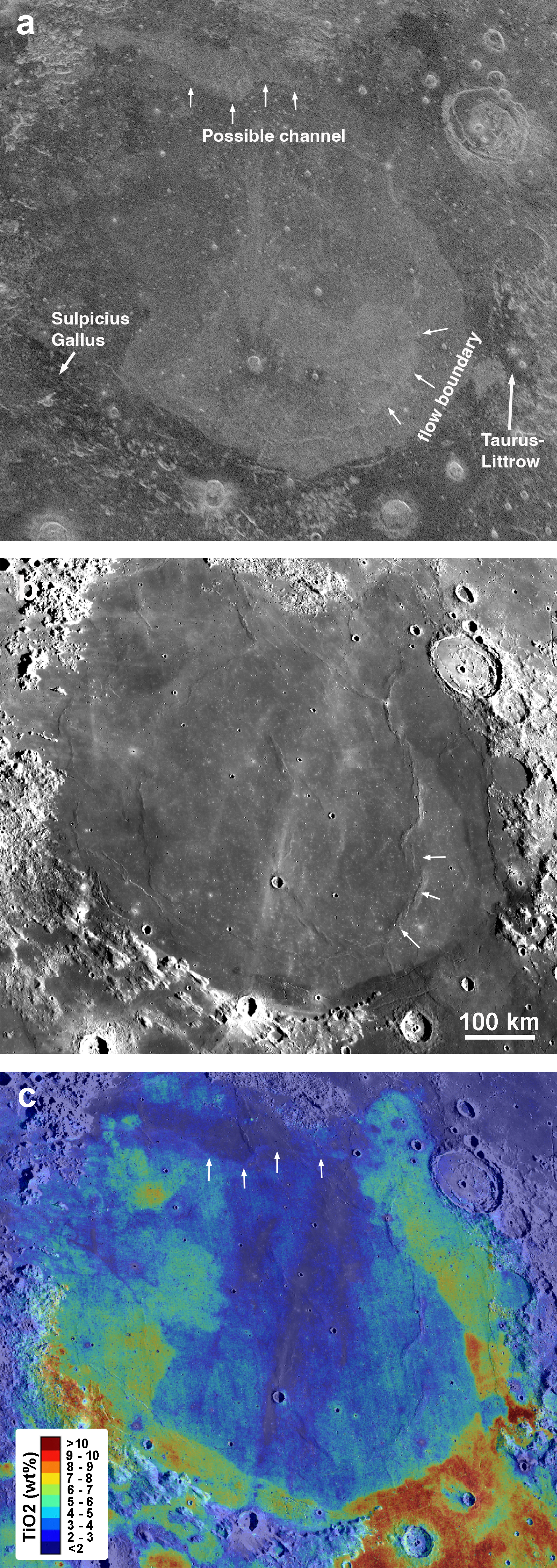}
    \caption{\label{fig:serenitatis} Arecibo P-band data reveal buried flows in Mare Serenitatis. (a) Arecibo P-band image of Mare Serenitatis; north is up. (b) Lunar Reconnaissance Orbiter (LRO) Wide Angle Camera image mosaic of the same area, created from images with large shadows to highlight topography. (c) Derived TiO$_2$ from LRO Wide Angle Camera data \citep{2017IcarusSato}. Arrows in (a) and (c) highlight a radar-dark, high-TiO$_2$ zig-zagging feature in the north. Arrows on the right in (b) indicate the rounded edge of a possible volcanic flow boundary, which is visible in radar (a) and partially visible in optical data (b). Two large pyroclastic deposits, Sulpicus Gallus and Taurus-Littrow, are also marked.  In areas with high TiO$_2$ (c), the radar wave is attenuated, and the radar backscatter power (a) is lower. In areas of low TiO$_2$, the radar wave is able to penetrate through the upper regolith and reveal the buried flow structure \citep{Campbell2014JGRE}.}
\end{figure}

The Arecibo data proved to be useful for studying volcanic regions of the Moon, particularly at P-band, where the radar penetration depth leads to dramatic views of the lava flows that built up the mare. Across Mare Serenitatis, the P-band data revealed radar-bright structures with apparent flow lobes and channels \citep{Campbell2009GeoRL, Campbell2014JGRE} (Figure~\ref{fig:serenitatis}). The spatial distribution of these features is inversely correlated with the ilmenite (TiO$_2$) content derived from spectral data. Ilmenite is common in some lunar basalts and has a high microwave loss tangent, leading to significant radar attenuation. Where the mare basalts have high TiO$_2$, the radar wave is attenuated and the radar backscatter is low. In places where the volcanic flows have low TiO$_2$, the regolith above those flows is similarly low in ilmenite, and the radar wave is able to penetrate and reveal buried rugged surfaces and structures \citep{Campbell2014JGRE}. Other lunar mare, including Imbrium, also have multiple flows visible in the radar data with backscatter properties that likely depend on the dielectric properties and emplacement of the flows \citep{ Morgan2016JGRE}. In the northern part of Mare Serenitatis, a large, radar-dark linear feature, invisible in optical images but apparent in TiO$_2$ maps (Figure \ref{fig:serenitatis}c), may be a buried volcanic vent or lava channel \citep{Campbell2014JGRE}. The P-band data also revealed mare deposits that have been buried by subsequent large impacts (“cryptomare”), such as those near the Orientale basin \citep{Campbell2005JGRE}.

Pyroclastic deposits, the fine glassy beads formed as a result of explosive volcanism on the Moon, are typically very dark in radar images owing to their smooth surface and relative lack of embedded rocks.  Arecibo polarimetric radar images made it possible to map the extent and properties of multiple large pyroclastic deposits, as well as possible pyroclastics and rugged flows associated with smaller volcanic domes \citep{Carter2009JGRE, 2009JGRECampbell}. Large pyroclastic deposits surround Mare Serenitatis, and a large radar-dark feature in the south that also has low CPR may be a previously unrecognized pyroclastic deposit or perhaps a lava flow with an unusual composition \citep{Carter2009JGRE}. The Aristarchus plateau has the largest extent of pyroclastic deposits. Radar data revealed a buried lava flow near the Vallis Schroterii rille and were used to map crater ejecta from Aristarchus and identify areas of deeper and less contaminated pyroclastics \citep{Campbell2008Geo}.

\subsection{Impact Craters}
\begin{figure}
    \includegraphics{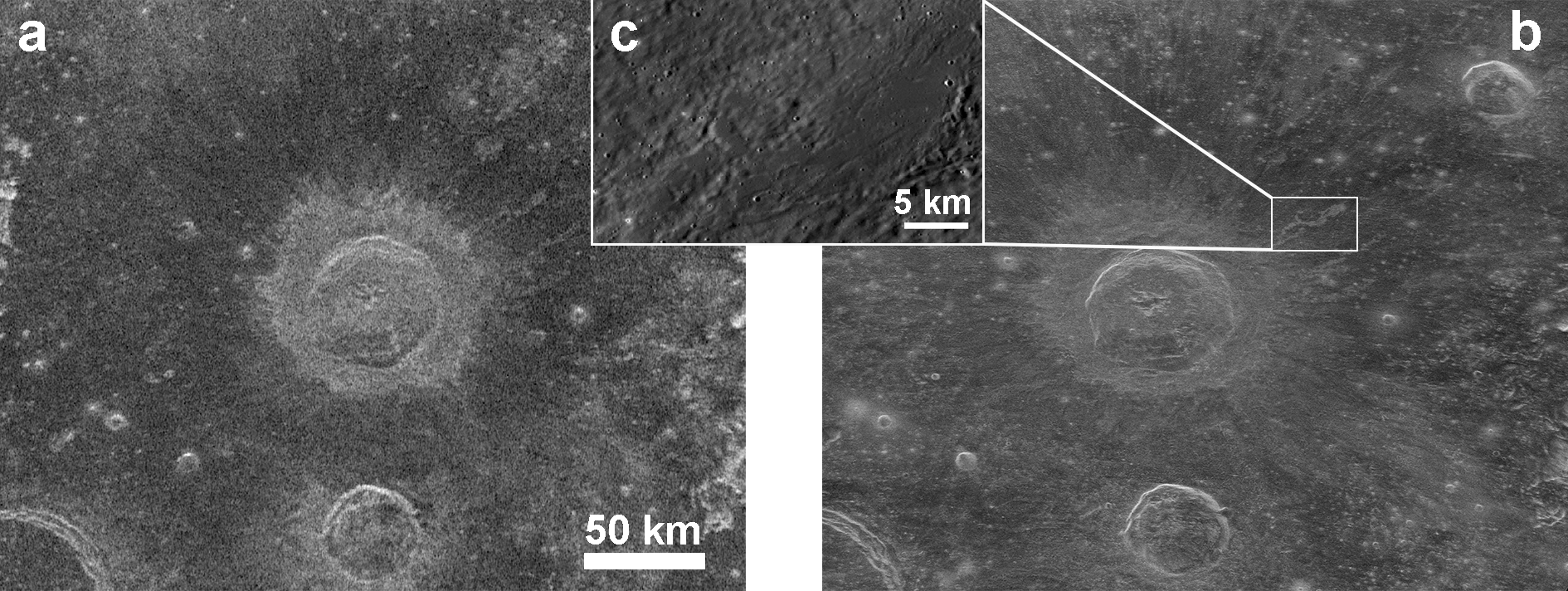}
    \caption{\label{fig:aristillus} The Aristillus crater has both a radar-dark halo and impact melt deposits. (a) P-band image at 400 m/pixel showing a bright ejecta blanket with impact melts to the northeast. A radar-dark halo extends out about 132 km from the crater \citep{Ghent2010Icarus} (b) An S-band image at 80 m/pixel shows the impact melts more clearly, as well as radar-bright streaks emanating from the crater that are caused by centimeter-sized ejected rocks. Because this is a fairly young crater, the dark halo is visible in both the P-band and S-band images. (c) The impact melts are visible as smooth deposits in LRO Wide Angle Camera images.}
\end{figure}

The Arecibo radar data also significantly improved our understanding of lunar impact craters and their extended ejecta deposits. Many lunar impact craters have radar-dark halos in the Arecibo P-band radar images \citep{Ghent2005JGRE} (Figure~\ref{fig:aristillus}). These dark regions are distal deposits that occur outside the brighter, rough, proximal ejecta blanket, and they also have low CPR values \citep{Ghent2005JGRE}. They are interpreted to comprise a meters-thick, rock-poor deposit that formed as part of the impact process \citep{Ghent2010Icarus}. Over time, the halos brighten as rocks are incorporated through regolith mixing and overturn \citep{Ghent2005JGRE}. For younger craters, the radar-dark halo is also visible in S-band data \citep{Ghent2005JGRE}. The oldest lunar craters ($>$3 Ga) do not have halos, demonstrating that these features evolve as the regolith ages \citep{Ghent2005JGRE, 2016IcarusGhent}.

At S-band, fresh impact craters are extremely radar-bright, and they have high CPR values that are indicative of rugged, rocky surfaces. Some craters have slightly buried radar-bright impact melt deposits that are visible in higher-resolution S-band images (Figure~\ref{fig:aristillus}) \citep{Wells2010JGRE, Campbell2010Icarus}. The relative paucity of radar-bright craters in the lunar highlands relative to the mare is evidence of a megaregolith 1 to 2 km deep that prevents rocky impact ejecta deposits from forming on craters smaller than 10-20 km because the surface is already fractured \citep{Thompson2009Geo}. Secondary craters and crater rays are also visible in the Arecibo S-band data and are often easier to detect and map than in optical data \citep{Wells2010JGRE, Wells2017Icarus}. Secondary craters can be identified in radar data by an asymmetry in their ejecta blankets, allowing them to be distinguished from small primary craters \citep{Wells2010JGRE}, which can improve age dating of lunar surfaces. Comparison of the S-band radar data with high-resolution optical images from the Lunar Reconnaissance Orbiter (LRO) Narrow Angle Camera showed that the radar-bright, high-CPR crater ray ``streaks” are likely formed via debris flow initiated by impacts that formed secondary craters \citep{Wells2017Icarus}. These small debris flows roughen the local surface and carry entrained blocks that lead to the radar-bright ray structures \citep{Wells2017Icarus}. Analysis of high-resolution S-band observations with the LRO's Mini-RF instrument revealed that the secondaries observed by \citet{Wells2010JGRE} are part of a South-pole crossing crater ray emanating from Tycho \citep{Rivera-Valentin.2024}. In the latter work, they further showed that the extended, asymmetric, radar-bright ejecta of secondary craters are more readily identified at large distances in full-Stokes polarimetric imagery compared to CPR alone. 

\subsection{Bistatic Observations with Lunar Reconnaissance Orbiter Mini-RF}
The LRO Mini-RF instrument was a synthetic aperture radar instrument operating at wavelengths of 12.6 and 4.2 cm. In 2010, the Mini-RF transmitter malfunctioned, and the instrument began to be used with the Arecibo S-band radar system in a bistatic configuration starting in 2012 \citep{Patterson2017Icarus} (later, other transmitters were used). For these observations, the LRO spacecraft was rotated so that the Mini-RF antenna pointed at the part of the surface illuminated by the Arecibo radar transmitter. These observations have a different viewing geometry than the typical monostatic radar systems, offering an opportunity to observe how the reflectivity and polarimetry change with differing scattering geometries. The ejecta of several lunar craters showed a possible opposition effect, where backscatter increases sharply at angles near nadir incidence \citep{Patterson2017Icarus}. An opposition effect was also observed across parts of the floor of the south polar crater Cabeus \citep{Patterson2017Icarus}. This response occurred in a part of the floor that is not in permanent shadow, but it may indicate the presence of water ice or a different type of near-surface structure than in some of the other observed craters.

\section{Mars}
Dual-polarization Arecibo S-band radar observations of Mars in the early 1980s revealed spatial variations in the depolarized echo component, i.e., in the same circular polarization as transmitted, in echo power Doppler spectra \citep{Harmon.1982, Harmon.1985}. The highest CPRs were largely associated with the major northern volcanic regions, suggesting roughness at the centimeter to decimeter scale. Comparisons between Arecibo S-band and Goldstone X-band observations further revealed that the decimeter-scale roughness was likely buried under a thin dust mantle \citep{Harmon.1992a}. Due to Mars’ rotation period, though, delay-Doppler radar imaging was precluded until the 1990 opposition, when the long code technique \citep{Harmon.2002} was used to mitigate ``over-spreading" (Doppler aliasing) of the signal \citep{Harmon.1992b}. Arecibo did not obtain the first radar images of Mars \citep{Muhleman.1991}, but, similar to Mercury observations, the post-upgrade Arecibo planetary radar system improved imaging resolution by an order of magnitude, leading to a spatially projected resolution of 3~km \citep{Harmon.2012, Harmon.2017}. S-band imagery continued to reveal the ancient Martian volcanic history. Lava flow channels long buried by dust were distinct in S-band imagery as radar-bright features. The highly reflective channels facilitated tracing them to source vents, as well as relative aging via superimposition. For example, a lava flow extending from Elysium Mons, through Marte Valles, and into Amazonis Planitia produced some of the brightest radar backscatter from Mars \citep{Harmon.1999, Harmon.2012}. This feature, though, is not fully expressed on the surface in topography or optical imagery, suggesting dust mantling. Such Arecibo radar imagery, along with orbital assets at Mars, have been used to constrain the age of the youngest lava flows, which may have formed in the late Amazonian Period \citep{Fuller.2002}. Furthermore, variation in the radar properties of the flows place constraints on their textures and thus their emplacement environment. \citet{Carol.2021} note that  the apparent high centimeter- to decimeter-scale roughness suggested by the circular polarization ratio combined with the meter-scale smoothness based on visible imagery indicate that the flow emplacement process may be akin to that of rubbly Icelandic p\={a}hoehoe flows and Peruvian blocky flows. 

Beyond characterization of Martian volcanism, Arecibo radar imagery has facilitated the robotic exploration of Mars \citep[e.g.,][]{Golombek.2003, Spohn.2022}. NASA’s InSight lander included the Heat Flow and Physical Properties Package, HP3, with the goal of measuring the geothermal heat flux at 3–5~m depth using a penetrator \citep{Spohn.2022}. Arecibo radar imagery was used during landing site selection to investigate the near-surface roughness properties in order assure sufficient radar reflectivity for the spacecraft's descent radar to mitigate the potential of encountering large buried rocks that would obstruct the penetrator experiment \citep{golombek_selection_2017,putzig_radar-derived_2017}. Although ultimately the experiment was only able to reach 37 cm due to a lack of friction, small rocks were not encountered in the top 30 cm (i.e., within $\sim$1/3rd of the radar penetration depth in S-band for dry planetary regoliths). Indeed, small rocks and debris were only encountered at depths greater than 31 cm, likely buried impact ejecta. 

The most favorable opposition for ground-based radar observations of Mars occurred in October 2020 because of the small Earth-Mars distance at that time. Arecibo observations of Mars, even with the Hurricane Maria–stricken system, would have resulted in much-improved SNR. Tragically, the suspended antenna structure suffered its first cable failure in August 2020 and later collapse in December, precluding the leveraging of this opportunity.

\section{Saturn System}
Just as the Arecibo telescope was becoming operational after commissioning in 2000, Saturn was at its most northern aspect. As a result, the round-trip light time was shorter than the Arecibo visibility window 
(135 vs.\ 165 minutes). The Saturn system was therefore observed at each opposition through 2008. It was also possible to receive the signals at the Green Bank Telescope, though the SNR was much lower, even though the latter could receive the full 165 minutes. Saturn itself does not reflect radio waves---they are simply absorbed by the atmosphere---but the rings and satellites do \citep{goldstein_radar_1973,ostro_radar_1980}.

\subsection{Titan}
\citet{campbell_radar_2003}'s observations of Titan showed a broad diffuse echo suggesting a moderately rough surface, but with a narrow peak at the center of the range of Doppler frequencies in a subset of observations. They interpreted these spiky echoes as indicative of liquid lakes on the surface. Cassini did later observe lakes, but not in the locations expected from the Arecibo data \citep{hayes_lakes_2016}. \citet{hofgartner_root_2020} compared the regions observed by both Arecibo and Cassini and interpreted the data as consistent with dried-up lakes, providing a very flat surface from ancient deposits. \citet{black_Titan_2011} compared the locations of the bright specular reflections in the Arecibo data to results from the Cassini 2.2-cm radar and found nothing special about those regions. As they acknowledged, the results remain somewhat mysterious.

\subsection{Saturn ring waves}
\begin{figure}
    \includegraphics[width=0.45\textwidth]{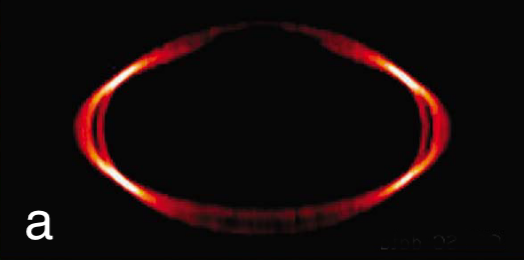}\includegraphics[width=0.45\textwidth]{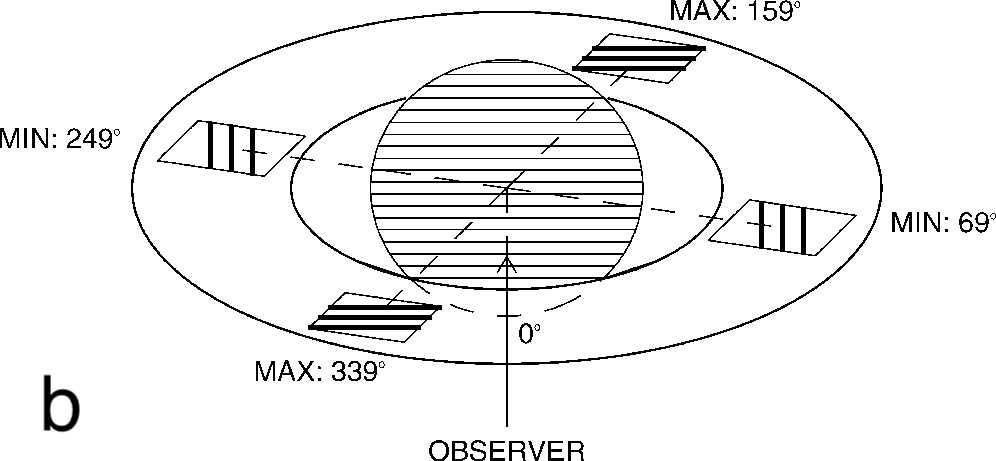}
    \caption{\label{fig:srings}On the left is an Arecibo radar delay-Doppler image of Saturn's rings from \citet{nicholson_radar_2005}, but reversed left-to-right from how it was shown in that work. The radar image has distance on the vertical axis and Doppler shift (that is, projected velocity) on the horizontal. The inner rings rotate faster than the outer ones, and thus the ring nearest the observer (near the bottom center of the image) ``crosses over" the inner ring as it moves towards the sides. Thus, the crossing points are brighter than the rest of the rings, as the signal is concentrated in a small region of the delay-Doppler image. The cartoon on the right from \citet{salo_photometric_2004} shows how the orientation of waves in Saturn's rings are predicted to vary with azimuth. The lines in the rectangular boxes show the predicted orientations of waves in the rings in those regions, where the waves are aligned with (top left and bottom right) and perpendicular to (bottom left and top right) the line of sight from Earth. These points are near the crossing points in the Arecibo image. The bottom-left and top-right regions are brighter than the other two, consistent with the prediction of wave orientation by \citet{salo_photometric_2004}. Without the waves, the radar image would be symmetric left-to-right}
\end{figure}

The rings of Saturn present a large surface area of solid material that scatters radar energy. These can be imaged much like a solid surface: The radar signal is separated into bins of distance from the observer by the difference in time delay between nearer and farther particles. The signal is further separated in Doppler shift, which depends on the orbital velocity of the particles. The radar images can be searched for out-of-ring-plane waves in the rings themselves, much like ocean waves: As viewed from shore, one sees the face of the wave, which presents a larger cross-section than if viewed from the side while swimming in the water. These waves were visible as higher brightness in some regions of the rings. Figure \ref{fig:srings} shows the observed ring images, along with a cartoon indicating where the rings are predicted to be brighter \citep{salo_photometric_2004}.
\newpage
\section{Near-Earth Objects}
In 1998, when the Arecibo planetary radar system came online after the Gregorian upgrade, most small asteroids, and in particular near-Earth asteroids, were understood to be collisional remnants of larger objects, but little was known about their structure. At this time, the LINEAR (Lincoln Laboratory Near-Earth Asteroid Search) program was regularly discovering near-Earth asteroids of 300$\,$m to 1$\,$km in diameter \citep{stokes_lincoln_2000}. The imaging resolution of the Arecibo radar was primarily limited by the rate at which data could be taken continuously, which was initially 5 MHz and soon 10 MHz, or 100 ns per sample. This yielded a range resolution of 15 m/pixel. 

A few near-Earth asteroids had known compositions from visible and near-infrared optical observations, and a small handful had been visited by spacecraft. These were all ``space potatoes”: irregular objects that could reasonably be collisional fragments of parent bodies in the main asteroid belt. This irregularity of shape was expected based on pre-upgrade Arecibo radar observations \citep{Ostro1986}.  It was also expected that ``small” asteroids, less than a few tens of kilometers in size, would be cratered monoliths \citep[e.g.][]{farinella_asteroids_1982}. The Galileo spacecraft flybys of 5-km asteroid (951) Gaspra and 15-km (243) Ida in the main belt \cite{sullivan_geology_1996,greenberg_collisional_1994} and the NEAR (Near Earth Asteroid Rendezvous) Shoemaker mission to near-Earth asteroid (433) Eros \citep{prockter_surface_2002} did not dispel this idea. 

However, in the intervening decades, we have learned that the near-Earth asteroid population, like the broader asteroid population, is much more heterogeneous than those initial observations suggested, with a variety of physical and spectral properties. The main contributors to the understanding of shape heterogeneity among near-Earth asteroids were the substantial increase in objects discovered by asteroid search programs \citep{stokes_lincoln_2000,larson_catalina_1998,pravdo_near-earth_1999,jedicke_observational_2002} and the imaging of objects with the Arecibo and Goldstone Radar Systems \citep{nolan_arecibo_2002}. These two systems were largely complementary; Arecibo provided much higher sensitivity, while Goldstone provided wider sky coverage, longer observation windows, and, by 2014, finer peak resolution for the brightest targets \citep{benner_HQ124}. 

\subsection{Asteroid Shapes and Configurations}
\label{sec:config}
\begin{figure}
\begin{center}
    \begin{minipage}{0.74\textwidth}\includegraphics[width=\textwidth]{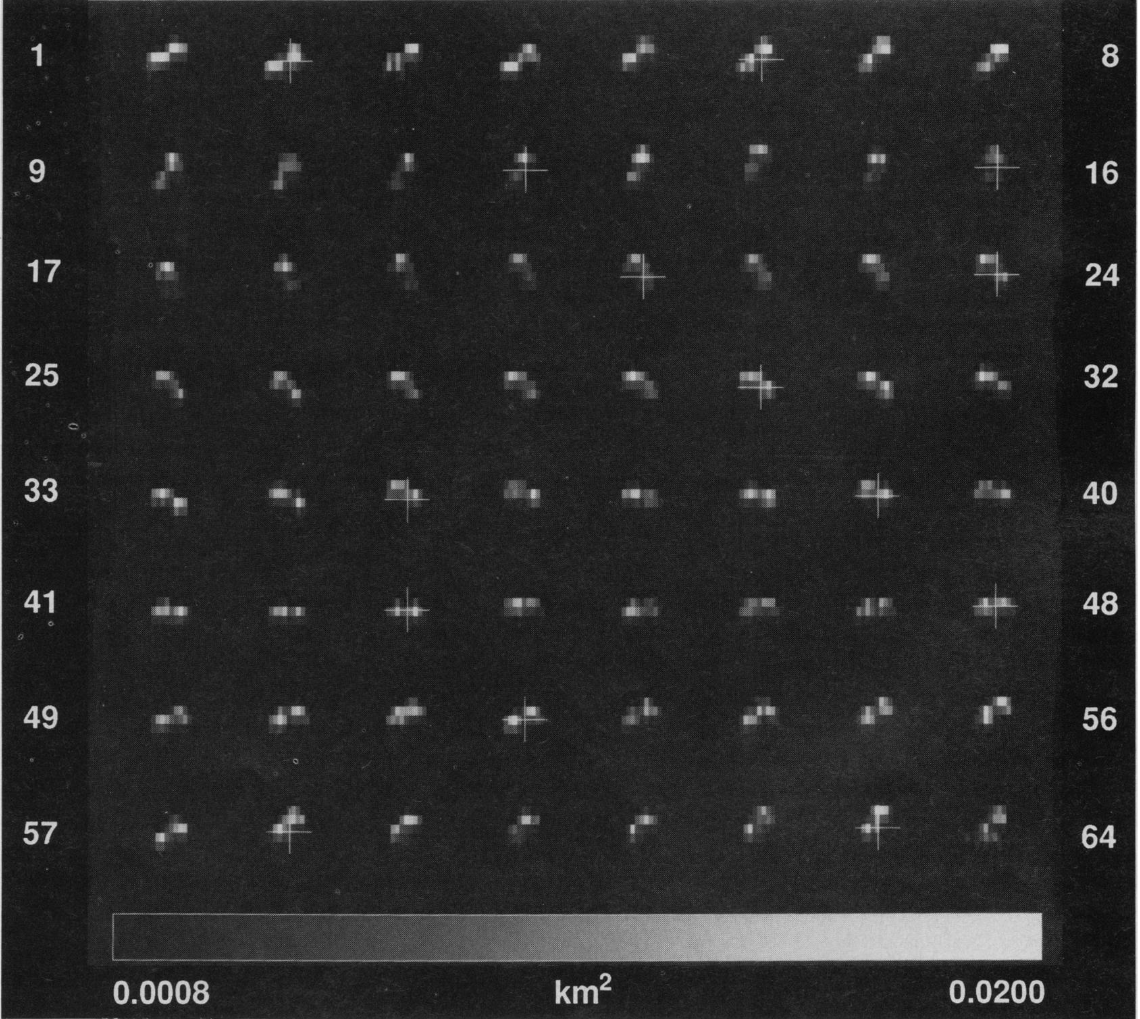}
    \end{minipage}
    \begin{minipage}{0.24\textwidth}
    \includegraphics[width=\textwidth]{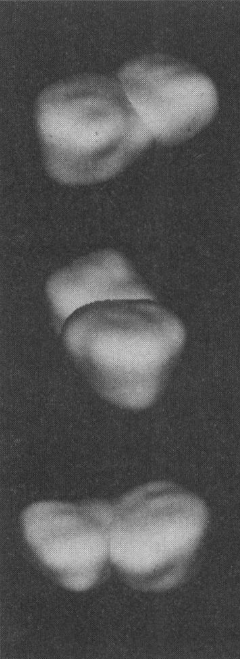}
    \end{minipage}
    \end{center}
    \caption{\label{fig:89PBdata}(Left) Radar images of near-Earth asteroid Castalia taken at the Arecibo Observatory on 1989 August 22, showing approximately half a rotation (from \citet{ostro_radar_1990}). Pixel spacing is 2 $\mu$s in the vertical (range) direction and 0.95 Hz in the horizontal (Doppler) direction, or 300 m $\times$ 150 m. This image sequence demonstrated the capability of planetary radar for the characterization of near-Earth asteroids, and may have helped persuade NASA to fund the upgrade to the radar system. (Right) Shape model derived from these data and observations performed on other days \citep{hudson_shape_1994}.}
\end{figure}

Although asteroids have been observed with radar since 1968\footnote{\citet{pettengill_icarus_1969} and \citet{goldstein_radar_1968} obtained weak detections of (1566) Icarus at Haystack and Goldstone, respectively, as it passed within 0.04 au of Earth in June of 1968.}, the close pass by Earth of newly discovered asteroid 1989 PB---later known as (4769) Castalia---allowed, for the first time, imaging (Fig. \ref{fig:89PBdata}) and shape modeling of what is now known as a ``potentially hazardous object", demonstrating the importance of radar observations. Castalia appeared to have an irregular bifurcated shape, though the resolution was low;  each of the lobes subtended two range pixels, so the surface details were not clear \citep{ostro_radar_1990}.

\begin{figure}
    \centering
    \begin{minipage}[b]{0.25 \textwidth}
    \includegraphics[width=\textwidth]{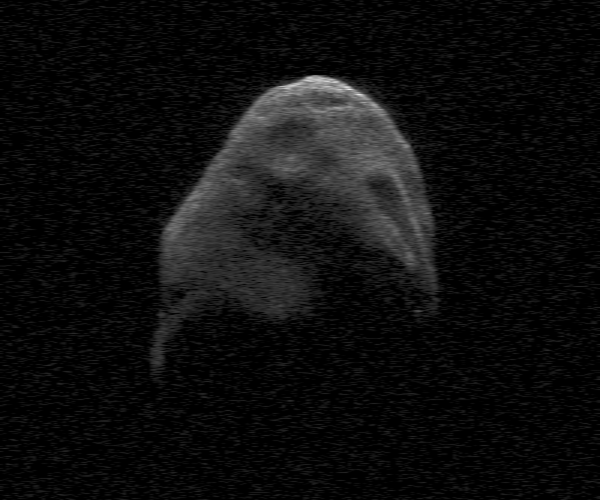}\\
    \includegraphics[width=\textwidth]{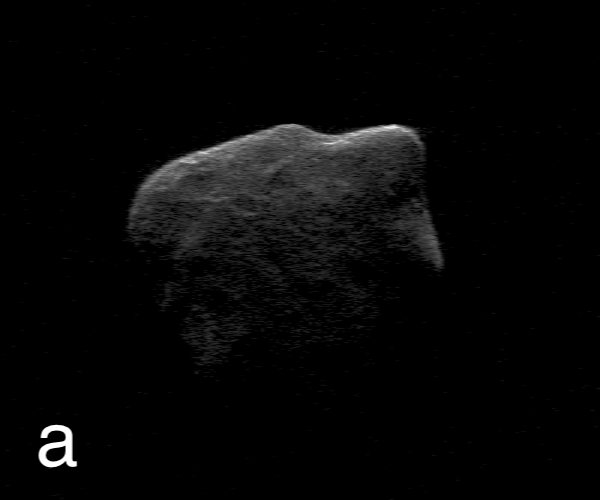}\\
    \end{minipage}
    \begin{minipage}[b]{0.20\textwidth}
\includegraphics[width=\linewidth]{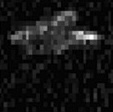}\\
\includegraphics[width=\linewidth]{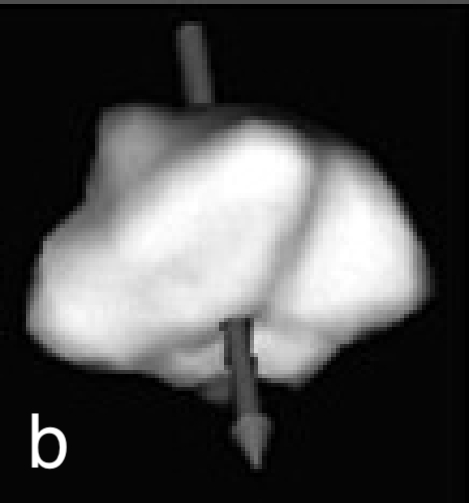}\\
    \end{minipage}
    \caption{Arecibo radar data of irregular near-Earth asteroids. (a) The two radar views of the $\sim7\,$-km-diameter asteroid 1999~JM$_8$ from 1999 August 1 and 5 appear remarkably different due to the variation in viewing geometry for this irregular, cratered object \citep{brozovic_JM8_2023}. There is no published shape model of this object. (b) Radar image (top) and shape model (bottom) of the $\sim$ 50-m-diameter asteroid (54509) YORP (from \citet{taylor_PH5}). The arrow through its center shows the rotation axis.
    }
    \label{fig:jm8}
\end{figure}

Other early targets were also irregular as expected, e.g., 1999~JM$_8$ (Figure~\ref{fig:jm8}a), (6489) Golevka (Figure~\ref{fig:jm8}c), and (4179) Toutatis. In addition, some objects were slowly tumbling, with rotation periods of tens of hours. Those facts combined to complicate the determination of their shapes, because they presented a different aspect each time they were observed but did not rotate enough in any given session to show how features were connected.

Then came observations of 1999~RQ$_{36}$---now known as (101955) Bennu---which was observed over several days in September 1999 and appeared to be round. These were quickly followed by observations of several other apparently spheroidal objects. In 2000, after many years of argument about the possibility of binary asteroids, and a few scattered detections in the main belt \citep{merline_asteroids_2002}, Arecibo acquired the first images of a near-Earth binary asteroid, 2000 DP$_{107}$.

\subsubsection{Bifurcated and Elongated Objects}
\begin{figure}
    \centering
    \begin{minipage}[b]{0.3 \textwidth}
    \includegraphics[width=\textwidth]{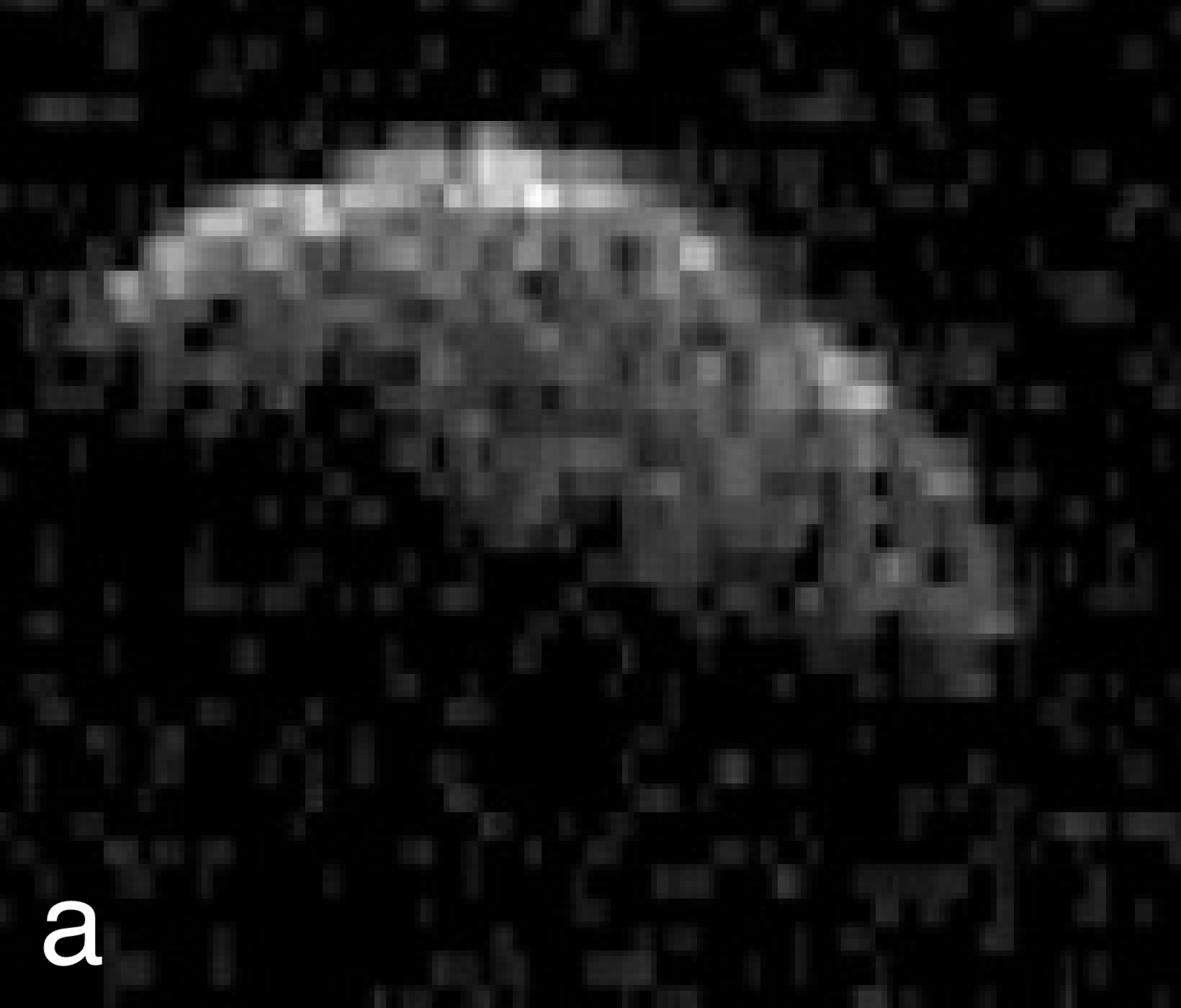}\\
    \includegraphics[width=\textwidth]{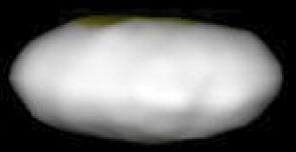}\\
    \end{minipage}
    \begin{minipage}[b]{0.31\textwidth}
\includegraphics[width=\linewidth]{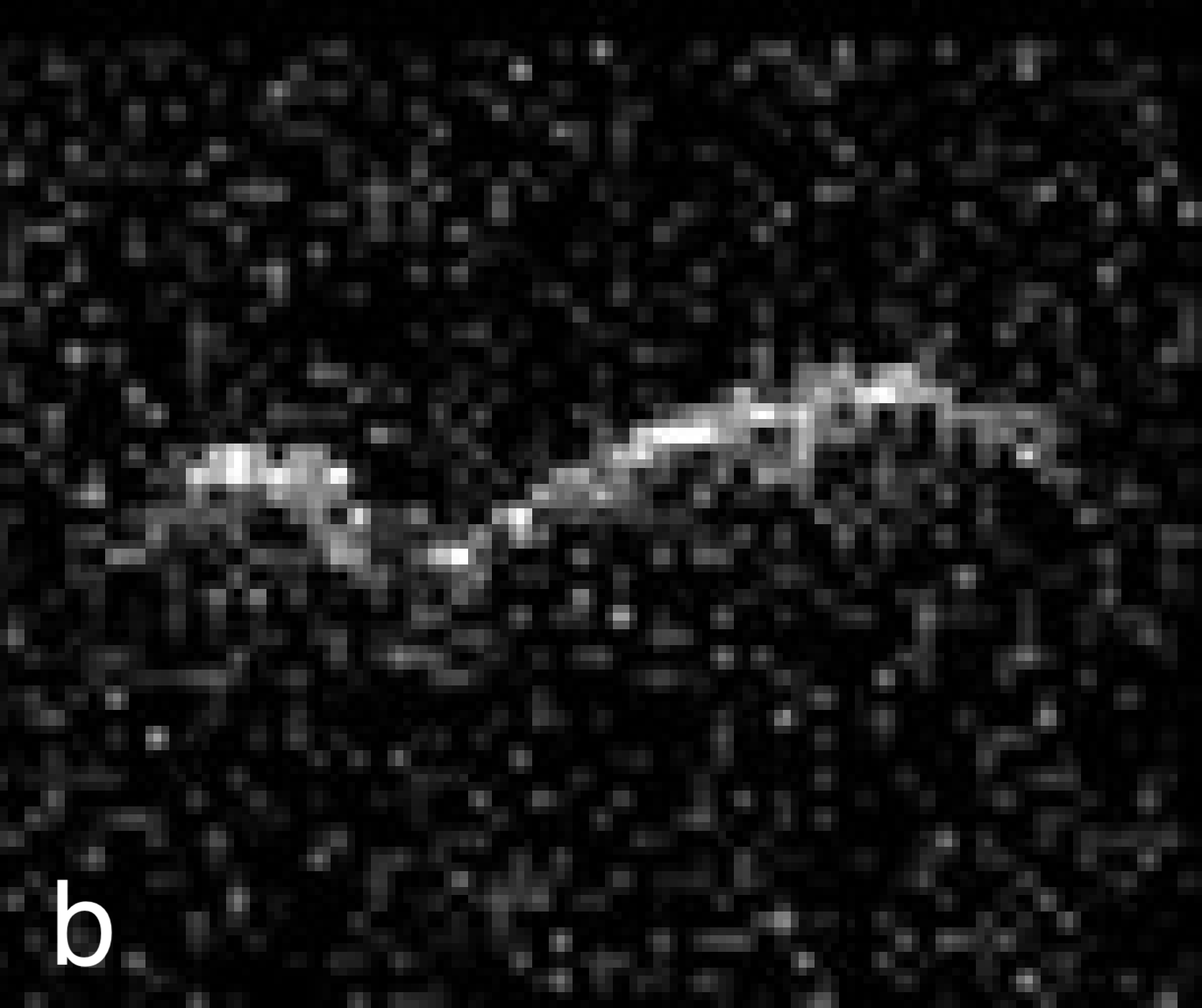}\\
\includegraphics[width=\linewidth]{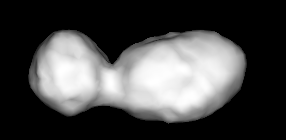}\\
    \end{minipage}
    \caption{Examples of elongated objects. (a) Radar image (top) and shape model (bottom) of (4660) Nereus, an oblong, non-bifurcated asteroid (from \citet{brozovic_Nereus_2009}). (b) Radar image (top) and shape model (bottom)  of 1996 HW$_1$, a bifurcated asteroid that resembles two objects resting on one another. Image rendered from the shape model of \citet{magri_radar_2011}.
    }
    \label{fig:HW1}
\end{figure}

In addition to Castalia, other bifurcated objects have been observed by Arecibo, with varying degrees of bifurcation. For instance, (8567) 1996~HW$_1$ appears remarkably like two ellipsoidal objects resting on each other \citep{magri_radar_2011}---a contact binary (Figure~\ref{fig:HW1}). Several formation mechanisms for contact binaries have been proposed, including rotational fission and reaccumulation from a catastrophic disruption of a larger parent \citep{meyer_strength_2024}. There are many other such objects in the Solar System at a wide variety of size scales, such as comets 8P/Tuttle and 103P/Hartley 2 and Kuiper Belt object (486958) Arrokoth, suggesting that there are likely multiple formation mechanisms and physical structures that are included in this category. There are also objects such as (4660) Nereus, which are elongated like 1996~HW$_1$, but show no bifurcation.

\subsubsection{Spinning-Top Spheroidal asteroids}
Approximately 1/3 of observed near-Earth objects between 100 m and 10 km in size are now known to be roughly spheroidal, and approximately half of those objects are part of multiple-asteroid (binary or triple) systems \citep{pravec_binary_2007}. This large abundance of spheroidal objects was unknown before 1998, because they are difficult to detect with lightcurve observations, and earlier radar observations were mostly of larger objects. Many of these objects have rotation rates very near the stability limit for a strengthless fluid body, suggesting that that their spheroidal shape is related to their rapid rotation. However, a few very slowly rotating spheroids have been observed, such as 1998~ML$_{14}$ \citep{ostro_ML14}, suggesting either that the rotation periods can evolve or that other mechanisms can produce a spheroidal shape.

Arecibo imagery of asteroid (66391) Moshup and its satellite Squannit, then known as 1999~KW$_4$, provided enough detail to construct a shape model \citep{ostro_KW4}.  The model revealed a ``spinning top” shape, which turned out to be characteristic of all of the spheroidal near-Earth asteroids observed with radar for which shapes have been modeled. This and other models of spinning-top asteroids from Arecibo radar data---e.g., 2003~SN$_{263}$ (Figure~\ref{fig:SN263}, right)---led to significant changes in our understanding of the evolution of small near-Earth asteroids and the forces that act on them. In particular, the idea that solar radiation forces, such as the Yarkovsky and YORP (Yarkovsky–O'Keefe–Radzievskii–Paddack) effects \citep{vokrouhlicky_yarkovsky_2015}, could alter these asteroids' orbits and spins, became widely accepted (see Section~\ref{sec:Yark}). The YORP effect, considered the most likely torque to act on small asteroids, can in principle both increase or decrease an object's rotation rate and may be responsible for the observed variation in rotation.
\subsubsection{Multiple-Asteroid Systems}
\begin{figure}
    \centering
    \begin{minipage}{0.79\textwidth}
       \includegraphics[width=\textwidth]{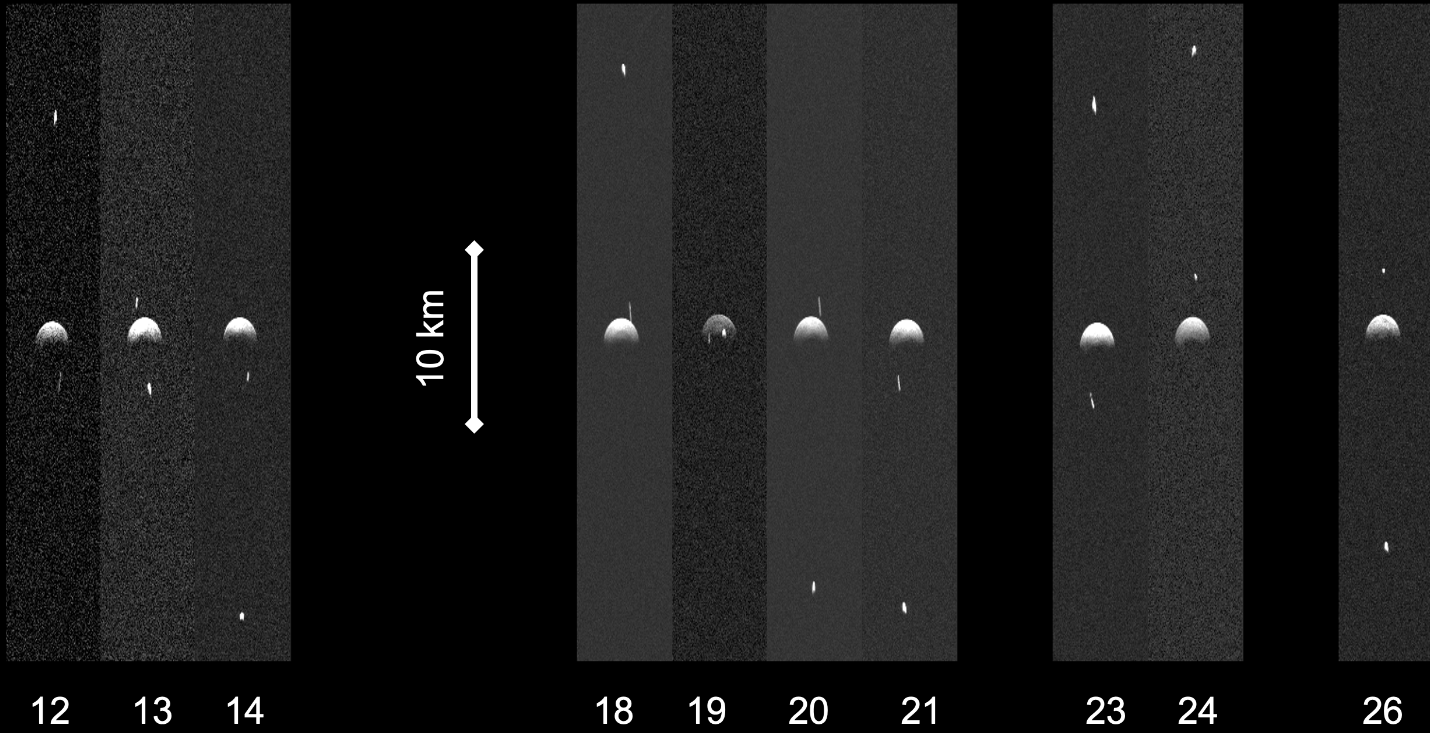}
       \end{minipage}
    \begin{minipage}{0.2\textwidth}
    \includegraphics[width=\textwidth]{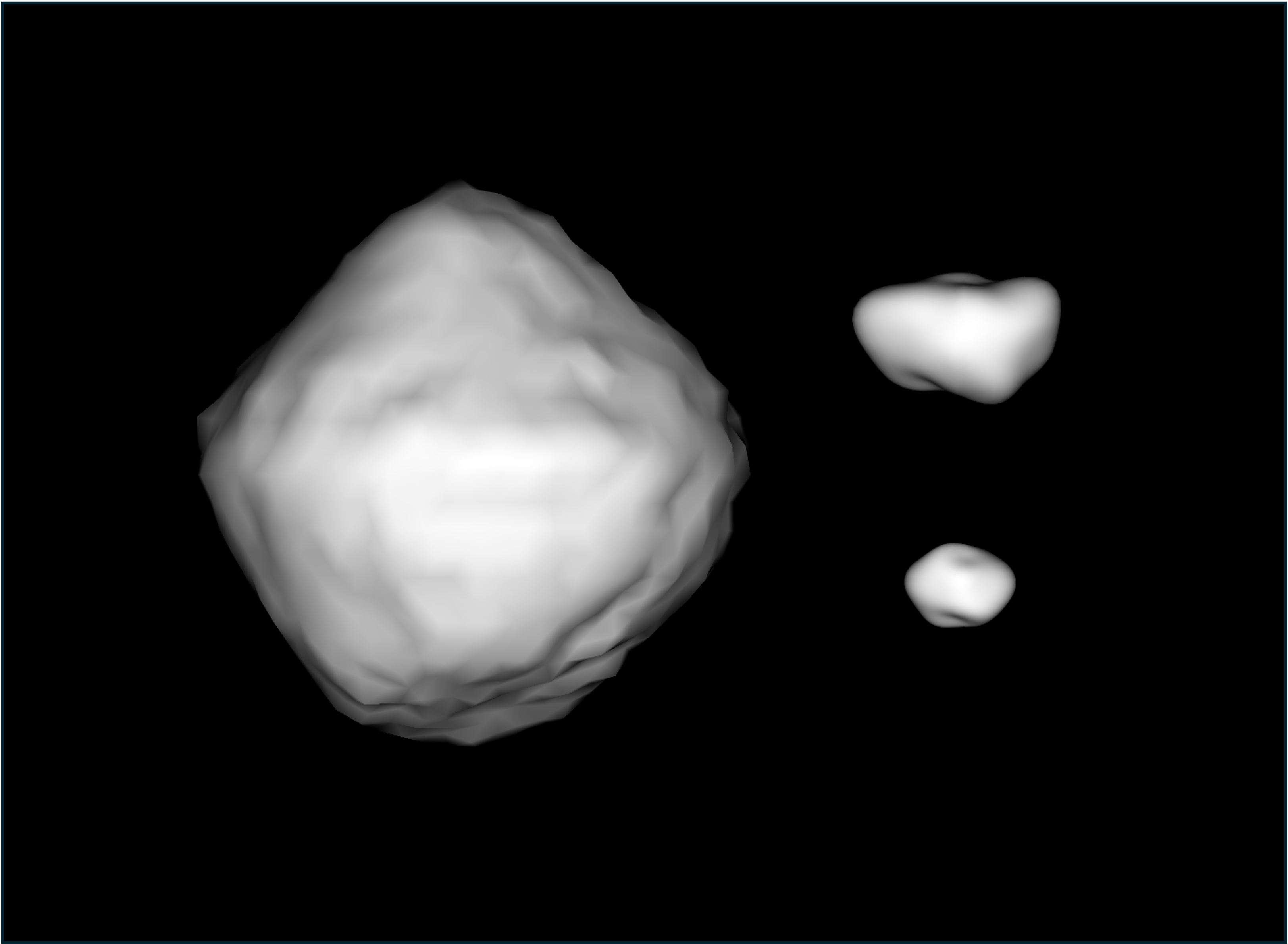}
    \end{minipage}
    \caption{(Left) Radar images of the triple asteroid system 2001~SN$_{263}$. Each panel in the image is from one day in February 2008, showing the primary asteroid and the two satellites (visible as white streaks). Blank spaces are days when no data were collected. 
    The larger, outer satellite orbits with a $\sim6\,$-day period, whereas the smaller, inner satellite orbits in $\sim 16.5\,$h. The inner satellite's rotation period appears to be the same as its orbital period---it is tidally locked---but the outer satellite rotates much faster than its orbit period. (Right) The derived shape model of the three components. The primary asteroid is  $\sim 2.8\,$km in diameter, and the satellites are $\sim 0.75$ and 0.5 km in mean diameter. After \citet{becker_SN263}
    \label{fig:SN263}}
\end{figure}

Shortly after the discovery of 2000 DP$_{107}$,  other objects including 2000~UG$_{11}$, 1999~KW$_4$, 1998~ST$_{27}$, and 2002~BM$_{26}$ \citep{margot_binary_2002} were identified as binary systems. Because radar can distinguish separation in range, it is possible to acquire unambiguous radar detections of multiple-asteroid systems, and unlike photometric detections, these do not depend on the orbital orientation of the objects \citep{pravec_binary_2007}. In 2008, Arecibo data led to the discovery that the asteroid 2001~SN$_{263}$ is in fact a triple asteroid system, the first to be observed in near-Earth space \citep{becker_SN263} (Figure~\ref{fig:SN263} and supplementary animation S1). Radar observations of multiple-asteroid systems are especially valuable because radar is sensitive to the absolute physical sizes of both the orbit and the objects, particularly orbital semimajor axis and asteroid volume, which enables estimation of the mass and density of the target---critical information for planetary defense \citep{NASEM_2023}

\subsubsection{Unclassified / Undetermined}
A significant fraction of spatially resolved objects remain unclassified. This is usually because only a single day of observations is available for an object that has not been well observed by other techniques, so that it is not possible to distinguish, for example, a contact binary viewed end-on from a spheroid or an elongated object if no change in shape is observed over the short observation window. The relatively short ($\leq 2.6$ h) observing windows at Arecibo contributed to this uncertainty, compared to Goldstone, which is both able (because of the flexible pointing) and required (because of the lower sensitivity) to observe for longer intervals. A number of asteroids could only be resolved into a few pixels, either because of Arecibo's range resolution limit of 7.5m or because of low SNR. We might again expect these to be cratered monoliths \citep[e.g.]{pravec_asteroid_2002}, though we have been surprised before.

No correlation has been identified between the shapes and the visible–infrared spectral types of asteroids, suggesting that, at these few-hundred-meter to few-kilometer scales, the mechanical properties do not depend heavily on composition. Binary systems have been found in nearly every spectral class, suggesting that mechanical rather than compositional properties influence their formation.

\subsection{Reflectivity properties}
\begin{figure}
    \centering
    \includegraphics[width=0.8\textwidth]{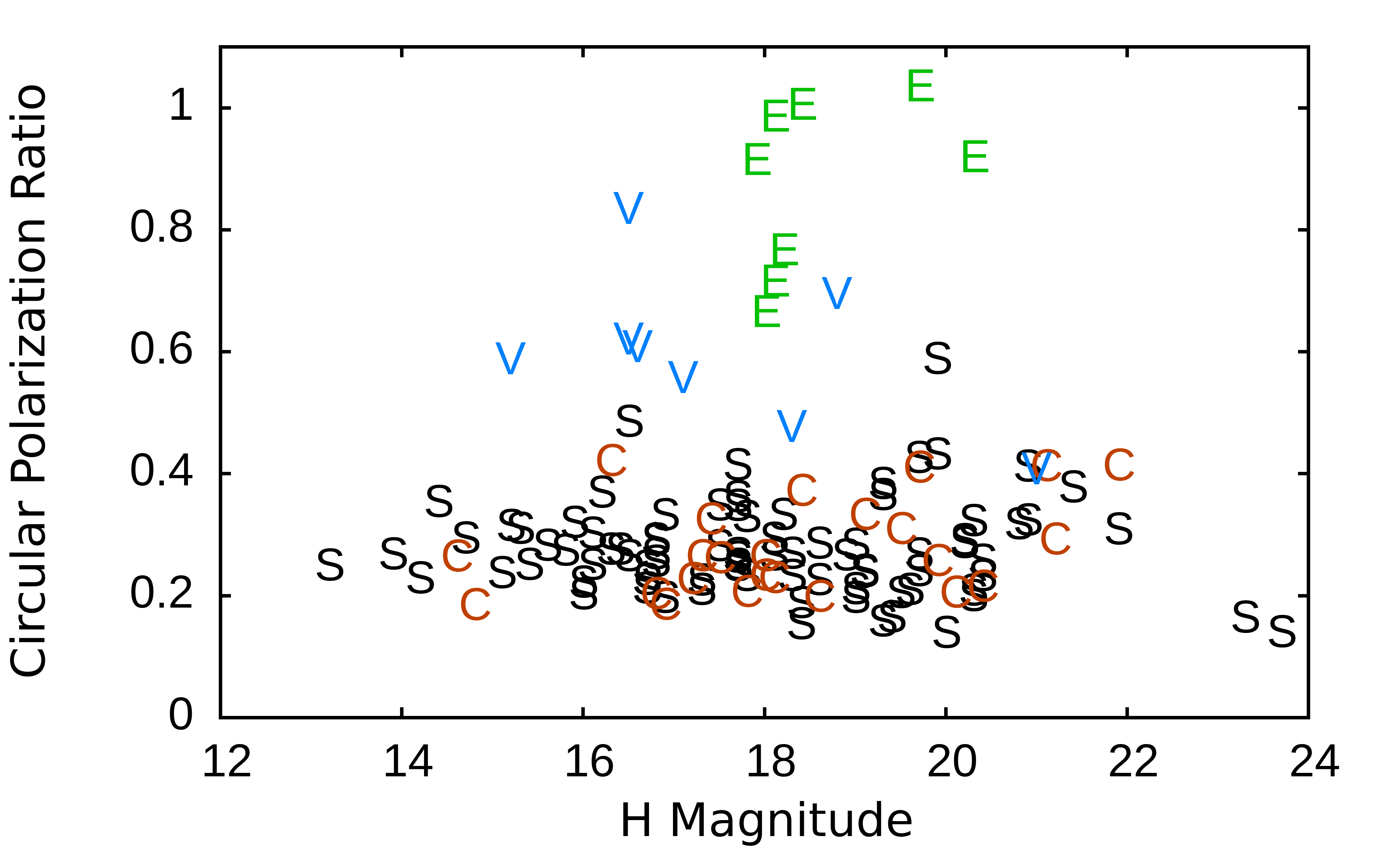}
    \caption{CPR as a function of absolute visual magnitude, a proxy for asteroid size because different spectral types have different albedoes. After \citet{benner_2008}. E-types can have polarization ratios as high as (or even exceeding) 1, V-types are clustered a bit over 0.5. No clear relationships between absolute magnitude and CPR are observed within the much more common S- and C-types.}
    \label{fig:SCOC}
\end{figure}

 The reflectivity properties of near-Earth asteroids are not well understood. Surprisingly, CPR has a specific correlation with V- and particularly E-class asteroids, as classified in the Tholen taxonomic system \citep{Tholen89}. As described in earlier sections, with the exception of ice, CPR is typically a structural, not a compositional effect. These two asteroid classes have larger CPRs, and thus different scattering properties, than the more common S- and C-type asteroids (Figure~\ref{fig:SCOC})\citep{benner_2008,rivera-valentin_radar_2024}. 

At least some M-class asteroids are believed to have metallic composition, and a number of these have high radar reflectivity as expected for metal \citep{shepard_radar_2015}. 

\subsection{The Yarkovsky and YORP Effects\label{sec:Yark}}
Because of its exceptional fractional precision---ranges can be measured to $<50$ m accuracy for objects millions of kilometers away---radar can be used to very precisely measure orbits of asteroids. By repeating such measurements, it is possible to measure the change in their orbits over time. The Yarkovsky effect is an acceleration due to thermal emission that was predicted theoretically and first measured for asteroids using Arecibo radar imaging of Golevka \citep{chesley_Yark}. This effect is responsible for slowly changing the orbits of asteroids smaller than about 40 km in diameter \citep{bottke_yarkovsky_2006}. 

The Yarkovsky effect has an important role in the delivery of meteorites to Earth and solved one of the long-standing problems in the velocity distribution of asteroid families. Asteroid families were discovered by \citet{hirayama_groups_1918}, who also suggested that they were likely the fragmented remains of a collisional disruption of a large body. However, the relative velocities of family members are too high to be collisional ejecta \citep[][and references therein]{nolan_impacts_2001}. The Yarkovsky effect changes the orbits of the fragments in a size-dependent way, so that the velocity distribution is different than the original distribution resulting from the collision. 
\citep{bottke_dynamical_2001,farinella_Yark_1998,nolan_impacts_2001}.

The Yarkovsky effect can also move asteroids into and out of resonances with the giant planets. These resonances can produce much more rapid changes in their orbits than the Yarkovsky effect, including delivering objects to near-Earth space or where they can be captured by the Earth. 

In addition, the Yarkovsky acceleration was used to estimate the mass and density of Bennu (see Section~\ref{sec:Bennu}).

Similarly, the YORP thermal emission effect alters the spin rates and pole orientations of small ($< \sim 40$) km asteroids \citep{scheeres_KW4,walsh_rubble_2018,bottke_yarkovsky_2006}. The YORP effect \citep{rubincam_YORP_2000} was first detected on asteroids using a combination of optical photometry and Arecibo radar observations \citep{taylor_PH5,lowry_direct_2007}. This effect, along with collisions, probably drives the evolution of small asteroids, potentially including resurfacing and satellite formation. 

\subsection{Spacecraft Targets}
\subsubsection{Itokawa}
\begin{figure}
    \includegraphics[width=0.7\textwidth]{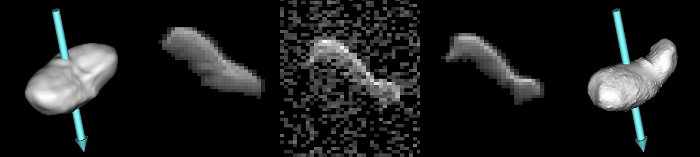}
    \caption{\label{fig:ito}Comparison of the Arecibo radar data of Itokawa with a radar-derived shape model \citep{ostro_itokawa} and the true shape as observed by the Hayabusa spacecraft \citep{gaskell_itokawa}. The columns show, from left to right: the radar shape model as it would appear from Earth with perfect vision; synthetic radar images based on that model; the actual radar data from 2004 June 19; synthetic radar images based on the true shape as determined by the Hayabusa mission; and the model of the true shape as viewed from Earth with perfect vision. The radar model gives a reasonable estimate of the asteroid's maximum size along each axis, but it does not capture the ``head and body" shape of the real object, largely due to conservative modeling assumptions intended to avoid over-interpreting noisy radar data.}
\end{figure}

Asteroid (25143) Itokawa was observed in preparation for the JAXA (Japan Aerospace Exploration Agency) Hayabusa mission, with the goals of readying the mission team for the encounter and providing an opportunity to validate radar imaging results \citep{ostro_itokawa}. This was the first attempt to use radar imaging to prepare for an asteroid encounter. The resulting shape model gave a reasonable prediction of the axis dimensions but was significantly fatter overall than the actual target and did not reflect the ``head and body'' nature of Itokawa (Figure \ref{fig:ito}). These differences were due to a combination of viewing geometries that did not cover the whole body and overly conservative assumptions about realistic asteroid shapes \citep{bramson_hayabusa_2009}.

\subsubsection{Bennu}\label{sec:Bennu}
Bennu’s orbit period is about two months longer than Earth’s, 437 days compared to 365.25, so it makes a relatively close pass by Earth every six years. After the initial observations discussed in Section \ref{sec:config}, observations were performed again in 2005 with the goal of capturing the maximum possible range of viewing geometries to determine whether Bennu was flattened (oblate) due to its rotation.

Shortly after the 2005 observations, Bennu was chosen as the target of the proposed OSIRIS asteroid sample return mission  \citep{drake_case_1987,drake_osiris-rex_2011,lauretta_osiris-rex_2017}, in part because the Arecibo radar data would provide useful information about it [C. Hergenrother, pers.\ comm.]. OSIRIS was not selected for launch in 2008, but in 2011, the somewhat larger-scale OSIRIS-REx mission became the third mission in the New Frontiers space flight program \citep{lauretta_asteroid_2010,lauretta_osiris-rex_2017}. By combining the 1999 and 2005 radar datasets with optical lightcurves, \citet{Nolan2013} produced a shape model of Bennu at approximately 20-m spatial resolution and determined the asteroid’s rotation rate and pole. These results were heavily used in mission design \citep{hergenrother_design_2014}.

Further radar observations determined the acceleration of Bennu due to the Yarkovsky force, which, combined with the measurement of the ``sail area” subject to solar radiation and the volume from the radar shape model, along with telescopic measurements of its thermal properties \citep{emery_thermal_2014}, allowed estimation of Bennu's mass with 12\% uncertainty \citep{chesley_orbit_2014}. The mass estimation dramatically simplified mission planning, as orbits could be computed in advance of arrival. The difference between the mass derived from radar and thermal telescope data and the mass eventually measured in orbit by OSIRIS-REx was 6\%, $0.5\sigma $ \citep{goossens_mass_2021, chesley_orbit_2014}.

\begin{figure}
\includegraphics[width=0.5\textwidth]{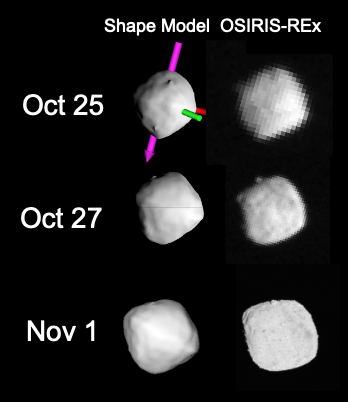}
\caption{\label{fig:BennuApp}The radar shape model of Bennu \citep{Nolan2013} (left) compared with actual images of Bennu acquired by the OSIRIS-REx spacecraft as it approached in late 2018 (right). The magenta arrow shows the model rotation axis and the red and green bars the coordinate axes. There are minor mismatches, but most of the features in the model were from real Bennu topography.}
\end{figure}

Figure~\ref{fig:BennuApp} shows some of the first resolved images of Bennu from OSIRIS-REx compared with predictions from the radar shape model. The rotation pole was off by approximately 5 degrees ($1.25 \sigma$), and Bennu is 1\% smaller in average diameter ($0.25 \sigma$) than the model, and the model features match remarkably well \citep{barnouin_shape_2019,daly_hemispherical_2020}. Even some of the unremarkable bumps on the shape model that might have been noise turned out to correspond to real boulders on Bennu.

\subsection{Comets}
\begin{figure}
\includegraphics[width=0.5\textwidth]{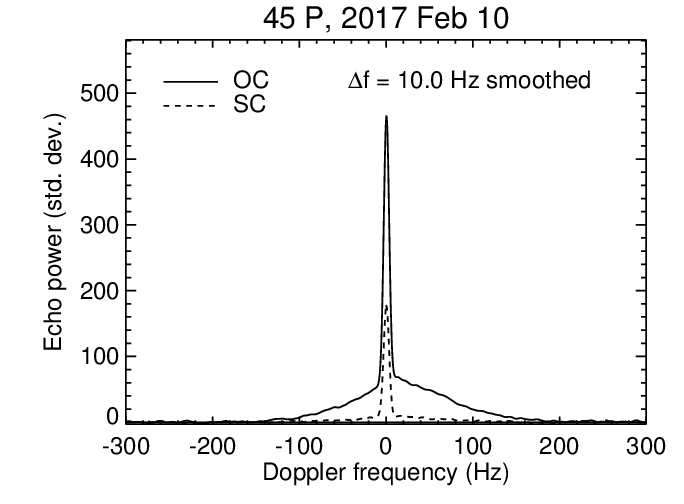}
    \caption{\label{fig:45P}This Doppler spectrum of comet 45 P/Honda–Mrkos–Pajdu{\v s}{\'a}kov{\'a} was obtained at Arecibo on 2017 February 10. It shows a  narrow component from the comet's nucleus, where the Doppler width is due to the object's rotation, flanked by a much broader ``skirt'' component from solid particles in the coma, where the Doppler spread is due to the motion of the particles. The asymmetry on the coma echo shows that the coma itself is asymmetric, with more particles moving towards the observer (positive Doppler) than away. The fact that there is some coma echo in the SC spectrum indicates that the particles are at least a few centimeters in size.}
\end{figure}

It is relatively rare that comets approach Earth closely enough that radar observations are possible, and thus the sensitivity of Arecibo has proven valuable. Beginning with comet 2P/Encke in 1980 \citep{kamoun_comet_1982}, 20 comets have been observed with the Arecibo radar. Of these, eight separate nuclei have been imaged (including two fragments of 73P/Schwassmann-Wachmann 3). Doppler spectra were obtained of all of these, some showing the nucleus, some showing a coma or ``skirt'' of large particles, or both.  Such particles were first detected in 1983 in comet C/1983 H1 IRAS-Araki-Alcock \citep{harmon_IAA}. Figure~\ref{fig:45P} shows a Doppler spectrum of C/45P Honda-Mrkos-Pajdu{\v s}{\'a}kov{\'a}, taken in 2017, one of the last Arecibo radar observations of a comet. The particles must be comparable in size to the radar wavelength to be visible. \citeauthor{harmon_CometsII}'s (\citeyear{harmon_CometsII}) model of particle velocities matches the data well, though fluxes and velocities vary somewhat from those obtained by other methods. Whether this difference is attributable to modeling approaches or the sizes of the particles is uncertain \citep{nolan_A2}. 

\begin{figure}
    \centering
    \includegraphics[width=0.2\textwidth]{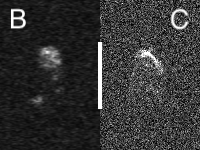}
    \caption{Radar image of 73P fragments B (left) and C (right). Fragment B appears to consist of several pieces, though this may simply be the result of an extremely irregular shape. Similarly, fragment C has a ``smear'' near the top that may be an irregular structure. White scale bar is 1 km.}
    \label{fig:73P}
\end{figure}

Comet 73P/Schwassmann-Wachmann 3 came close enough to Earth in 2006 that several fragments were imaged. The images suggested that large (greater than meter scale) pieces may have been leaving the surface and then fragmenting (Figure~\ref{fig:73P}) \citep{howell_radar_2007}.

Our understanding of comet surface processes is woefully incomplete. Radar observations are one of the few techniques to measure centimeter-scale grains, but the Goldstone system does not have the sensitivity to image any comet nuclei that are currently known.

\subsection{Heterogeneity of the Near-Earth Asteroid Population}
\begin{figure}
    \centering
    \includegraphics[width=0.7\textwidth]{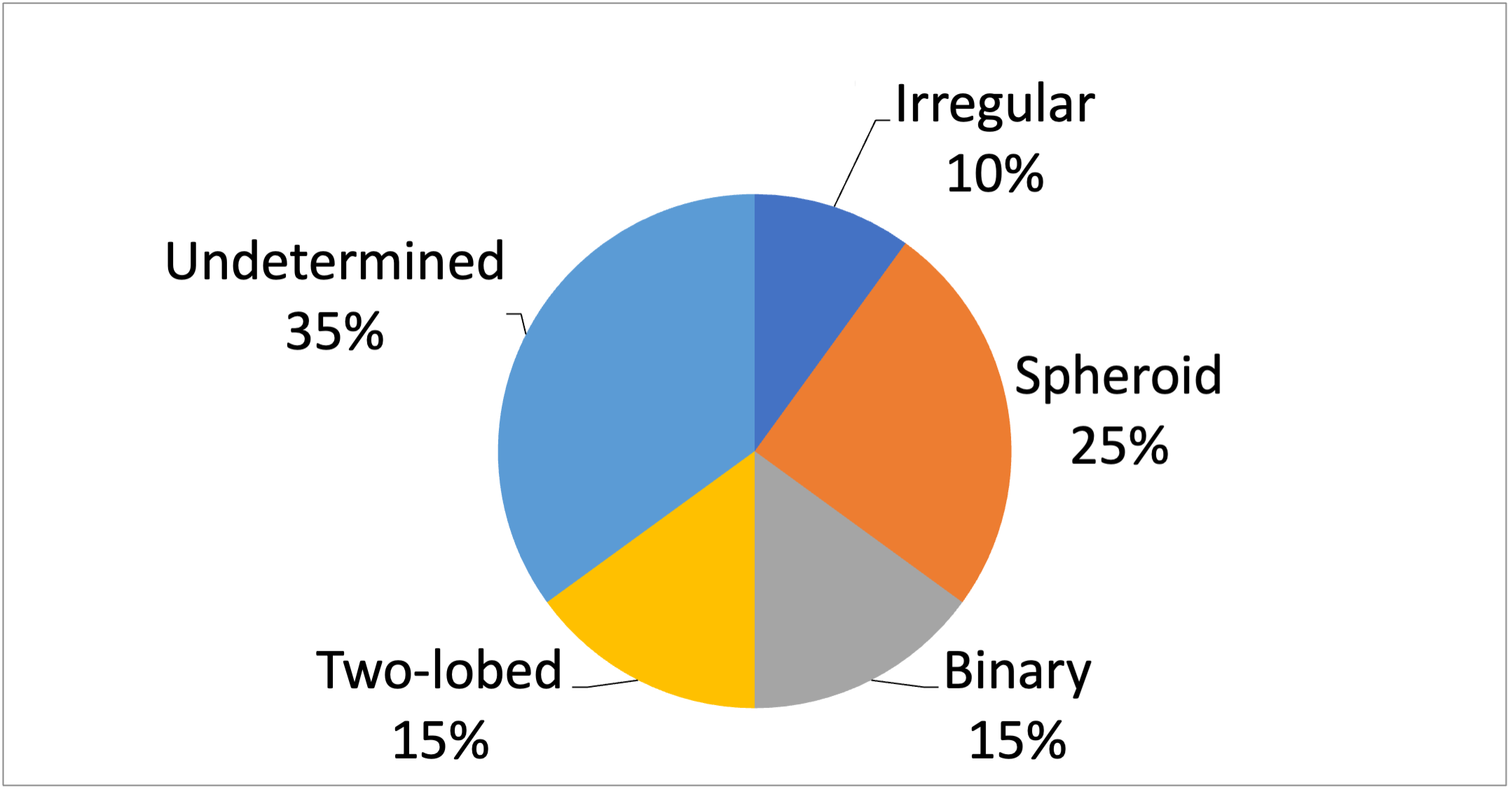}
    \caption{Fraction of radar-observed asteroids of a given shape or configuration. Objects can be Undetermined for several reasons, including insufficient observations to see the whole shape or rotating too slowly to be able to resolve in Doppler shift. The Undetermined category includes objects too small to be well-imaged, which may include a larger fraction of irregular objects. Statistics are from 400 well-observed objects as of 2019. Figure courtesy E. S. Howell.}
    \label{fig:class}
\end{figure}

The main result of 20 years of near-Earth object radar imaging at Arecibo is the observed diversity of structural properties. Figure~\ref{fig:class} shows the distribution of shape and configuration categories described above. ``Irregular'' asteroids, likely the most expected category before the discoveries described in this paper, account for a modest fraction of the total. Several different formation mechanisms have been proposed for both spinning-top and bifurcated objects, and it is plausible that multiple formation mechanisms are active in producing similar-looking objects.

\begin{figure}
    \centering
    \includegraphics[width=0.6\textwidth]{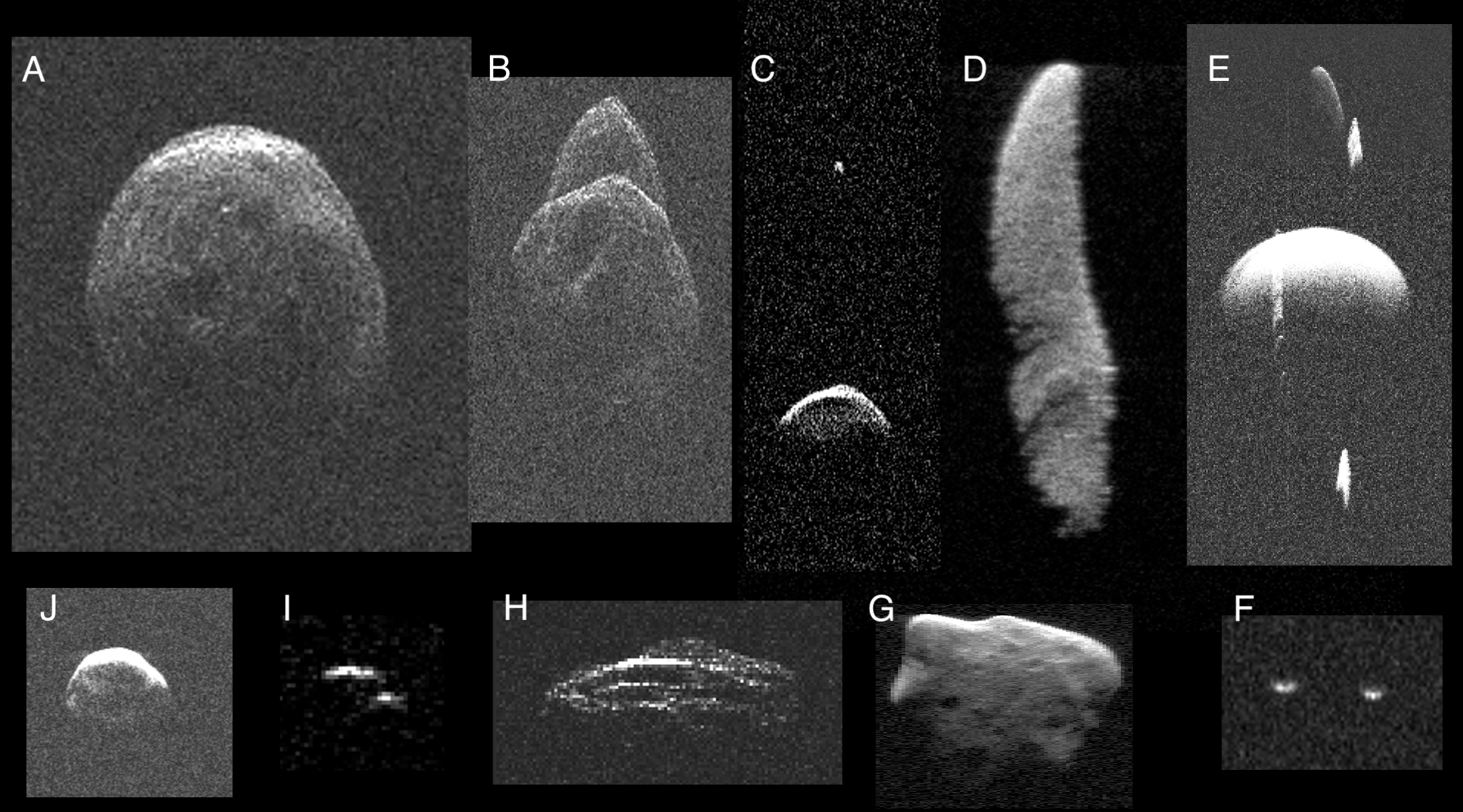}
    \caption{Rogues gallery of a few of the over 400 near-Earth asteroids imaged at Arecibo between 1997 and the collapse of the telescope in 2020. (A) (136849) 1998 CS$_1$, a $\sim1$-km spinning top with a 2.8 hour rotation period has clearly visible craters and boulders \citep{benner_CS1}. (B) (192642) 1999 RD$_{32}$, a $\sim$ 5-km contact binary with a 26 hour rotation period has angular-appearing segments \citep{taylor_RD32}. C) (164121) 2003~YT$_1$, a 1-km spheroid with a secondary asteroid with a 2.3 hour rotation period
    \citep{nolan_YT1}. Supplementary animation S2 shows some of the radar data showing the motion of the secondary. (D) (163899) 2003~SD$_{220}$, the most potato-like irregular object observed at Arecibo. It is 2 km long and rotates in $\sim$ 12 days \citep{rivera-valentin_SD220}. (E) Sum over four days of the 1.5-km spinning-top (66391) Moshup (1999~KW$_4$) showing the 500-m satellite Squannit as it orbits. Moshup rotates in 2.7 hours and Squannit in 17 h, equal to its orbit period. \citep{ostro_KW4} (F) Symmetric binary system (190166) 2005~UP$_{156}$ with two $\sim$ 900 m components orbiting a common center of mass, tidally locked so that both objects rotate at the 40.5-hour orbital period and each object always sees the same face of the other \citep{taylor_UP156}. (G) (53319) 1999~JM$_8$, a 7-km irregular object that is tumbling with an approximately 7 day rotation. No spheroidal objects this large have been discovered, so irregular shapes may be more common in this size range \citep{brozovic_JM8_2023}. (H) (436724) 2011~UW$_{158}$ is a rapidly rotating ($\sim35$ minute period) $300\times600$ meter ``space raisin'' based on its appearance in this image \citep{naidu_UW158}. (I) (138175) 2000~EE$_{104}$, a slowly rotating (14-hour period) 250-m probable contact binary \citep{howell_EE104}. (J) (29075) 1950~DA is a $\sim 1$-km spheroid with a 2.1 hour rotation period that has a small chance (0.04\% at the time of writing) of impacting Earth in the year 2880, the highest of any currently known object \citep{giorgini_1950DA,farnocchia_1950DA,chesley_DA}. }
    \label{fig:gallery}
\end{figure}

Figure~\ref{fig:gallery} illustrates a wide variety of shapes and textures. This diversity implies a variety of mechanisms of formation and possibly of delivery from the main belt to near-Earth space. In addition, though it is often assumed that meteorites are a reasonably complete sample of the asteroid population, there are likely dramatic differences in the meteorite delivery efficiency of these different objects due to differences in ability to survive atmospheric entry due to material properties \citep[e.g.,][]{orgel_evaluating_1998,demeo_compositional_2015} and entry velocity due to differing orbits \citep[e.g.,][]{binzel_compositional_2019,granvik_identification_2018}. This structural heterogeneity, not clearly associated with material composition, can only add to differing delivery efficiencies, for example, highly-fragmented asteroids may be less likely to survive atmospheric entry to produce meteorites than monolithic ones, or alternatively, may be tidally disrupted by Earth encounters resulting in increased cross-section for capture.

Thanks to its sensitivity, Arecibo was capable of observing approximately twice as many objects as its nearest competitor \citep{naidu_capabilities_2016}, and over its lifetime it observed approximately 900 individual near-Earth asteroids and comets, more than 100 per year at its peak, compared to the 51 total (including 32 at Arecibo) observed anywhere before the Gregorian upgrade in the late 1990s \citep{zambrano-marin_radar_2023}. 

\section{Closing Remarks}
This is where one customarily points to the future of the field and discoveries to be made. However, the Arecibo telescope collapsed on December 1, 2020, and is not likely to be rebuilt \citep{lloreda_Shutdown}. No existing or planned facility can match its sensitivity, flexibility, or number of observable asteroids \citep{Marshall2023}. No other ground-based system will observe the subsurface of Mars or Venus, or provide new data of Mercury or the icy moons of Jupiter and Saturn. As of 2025, only spacecraft missions, each one comparable in cost to building the Arecibo Observatory and operating it for 60 years, can achieve such high sensitivity. With the loss of Arecibo just as the next generation of asteroid discovery programs \citep[the Rubin Observatory and the NEO Surveyor mission,][]{sonnett_neo_2021} is expected to ramp up, this window into the formation and evolution of Solar System objects will darken.


\section*{Data Availability}
Many of the Arecibo data sets are publicly available through the NASA Planetary Data System. These include the lunar S-band and P-band data sets \citep{Campbell2011pdsdata, Campbell2007pdsdata} available at the PDS Geosciences node (\url{https://pds-geosciences.wustl.edu/missions/lunar_radar/index.htm}, \url{https://pds-geosciences.wustl.edu/missions/sband/index.htm}). Much of the processed Venus data \citep{Campbell2021pdsdata} is available at 
\url{https://pds-geosciences.wustl.edu/missions/venus_radar/index.htm}.
There are other efforts in progress to deliver the asteroid and comet data to the PDS Small Bodies Node, with an initial set accepted \citep{nolan_arecibo_2024}.

\section*{Acknowledgments}
This work was funded by the National Aeronautics and Space Administration (NASA) through the Solar System Observations program under grant Nos. NNX10AP64G, NNX12AF24G, and 80NSSC19K0523, and the Discovery Data Analysis Program under grant No. 80NSSC24K0062, and by the National Science Foundation. The Arecibo Observatory was part of the National Astronomy and Ionosphere center, operated by Cornell University (until 2011), SRI International (from 2011 to 2018), and the University of Central Florida (from 2018 to 2023) under cooperative agreements with the National Science Foundation and with support from NASA

\newpage

\bibliography{RMPRadar}
\end{document}